\documentclass[preprintnumbers,amsmath,amssymb,10pt,twocolumn,a4paper]{revtex4}
\usepackage{geometry}
\usepackage{mathtools}
\usepackage{xcolor}
\usepackage{graphicx}
\usepackage{dcolumn}
\usepackage{bm}
\newcommand{\be}{\begin{equation}}
\newcommand{\ee}{\end{equation}}
\newcommand{\bd}{\begin{displaymath}}
\newcommand{\ed}{\end{displaymath}}
\newcommand{\BE}{\begin{eqnarray}}
\newcommand{\EE}{\end{eqnarray}}

\newcommand{\bn}{\ensuremath{\mathbf{n}}}

\newcommand{\kk}{^{(k)}}

\begin{document}

\title{Zealots in multi-state noisy voter models}
 
\author{Nagi Khalil}
\affiliation{Escuela Superior de Ciencias Experimentales y Tecnolog\'ia (ESCET) \& GISC, Universidad Rey Juan Carlos, M\'ostoles 28933, Madrid, Spain}
\email{nagi.khalil@urjc.es}

\author{Tobias Galla}
\affiliation{Instituto de  F\'iısica  Interdisciplinar y Sistemas  Complejos, IFISC (CSIC-UIB), Campus Universitat Illes Balears,  E-07122 Palma de Mallorca, Spain}

\affiliation{Department of Physics and Astronomy, School of Natural Sciences, The University of Manchester, Manchester M13 9PL, UK}
\email{tobias.galla@ifisc.uib-csic.es}

\date{\today}

\begin{abstract}
The noisy voter model is a stylised representation of opinion dynamics. Individuals copy opinions from other individuals, and are subject to spontaneous state changes. In the case of two opinion states this model is known to have a noise-driven transition between a unimodal phase, in which both opinions are present, and a bimodal phase in which one of the opinions dominates. The presence of zealots can remove the unimodal and bimodal phases in the model with two opinion states. Here, we study the effects of zealots in noisy voter models with $M>2$ opinion states on complete interaction graphs. We find that the phase behaviour diversifies, with up to six possible qualitatively different types of stationary states. The presence of zealots removes some of these phases, but not all. We analyse situations in which zealots affect the entire population, or only a fraction of agents, and show that this situation corresponds to a single-community model with a fractional number of zealots, further enriching the phase diagram. Our study is conducted analytically based on effective birth-death dynamics for the number of individuals holding a given  opinion. Results are confirmed in numerical simulations.
\end{abstract}
\maketitle

\section{Introduction}
\label{introduction}

Three main components of individual-based models of natural and social processes are the states the individual constituents can be in, the topology of interactions, and the dynamics of these interactions. The term `states' refers to dynamic properties agents can hold, for example they may be infected or recovered in a model of an epidemic, or be of a particular opinion in a model of social dynamics. The interaction network describes who a given individual can interact with, and the dynamical rules specify the details of the interaction process (for example infection, or adopting an opinion held by another agent) \cite{cafolo09,sech14}. It is well established that the details of state space, topology, and interaction rules have significant consequences on the global behaviour emerging in interacting-agent systems \cite{cafolo09,krb_book}.

The so-called voter model (VM) is a good illustration of this. The VM provides a stylised description of the dynamics of opinions in a population of voters \cite{clsu73,holi75}. In the most basic version, each individual holds one of two possible opinions. An individual's state can change by means of an imitation process. More precisely the individual copies the opinion state of one of its neighbours on the interaction graph. This dynamics comes to a halt when consensus on one opinion is reached. If the effective dimension of the network is below two, populations described by the VM will always evolve towards consensus. In higher dimensions multiple opinions can co-exist indefinitely \cite{befrkr96,suegsa05,sore05}. 

The VM is not only sensitive to the topology (dimension) of the interaction network, but also to the inclusion of additional randomness in the dynamics. This is very apparent in the context of the so-called `noisy voter model' (NVM). In this model individuals interact via the above imitation process, but they can also change opinion spontaneously without interaction with anyone else in the population \cite{figuzi89,ki93,grma95}. One main consequence of this modification is the removal of consensus states as absorbing endpoints of the dynamics. We note that care needs to be taken when interpreting the word `noisy' in `noisy voter model'. The standard (`non-noisy') VM contains an element of stochasticity as well: At any iteration an agent is chosen at random for update, and then copies the state of a randomly chosen neighbour. The term 'noisy' in NVM is used to indicate the possibility of spontaenous state changes of the individuals.

One main object of interest in the NVM with two opinion states (labelled $1$ and $2$) is the stationary distribution $P^{\rm st}(n_1)$, for the number of agents $n_1$ holding opinion $i=1$. The number of individuals who are in opinion state $i=2$ is $n_2=N-n_1$, if $N$ is the total size of the population. Assuming a non-zero noise strength, the support of this distribution is $0\leq n_1\leq N$. If the noise is sufficiently small, the system spends most of its time close to the consensus states ($n_1=0$ and $n_1=N$, respectively), and travels from one consensus state to the other. The system is said to be in the `bimodal' state: $P^{\rm s}(n_1)$ has bimodal shape, with peaks at $n_1=0$ and $n_1=N$. On the other hand, if the population is finite and the noise strength is above a certain threshold value, the system is in the so-called `unimodal phase'. Assuming that there is no intrinsic preference for any of the opinion states, the stationary distribution has one single maximum at $n_1=N/2$. The topology of the interaction networks does not affect the nature of this transition. It only acts to modify the threshold value of the noise amplitude separating the two phases \cite{catosa15,catosa16,pecasato18}. 

Owing to its simplicity and analytical tractability, the NVM has been studied and generalized  \cite{nvm1,nvm2,nvm3,nvm4,nvm5,nvm6,nvm7,nvm8,copest16,arkhtosa18,ko19,ko20,ju05} in various different directions. This includes non-linearity in the imitation rates \cite{pecasato18b,jesz19}, memory effects \cite{daca07,arpetorasa18,pekhto19,pekhto20}, the introduction of contrarians \cite{khto19} or zealots \cite{nagi}, and multi-state noisy voter models  \cite{francisco}.

Multi-state VM are variants of the VM in which each individual holds one of $M\ge 2$ opinions. One main focus of the analysis of multi-state VM without spontaneously opinion-changes has been the time it takes to reach consensus (the so-called consensus time) \cite{ta84,co89,bablmc07}. Other works are concerned with the geometry of the ordering process by which the system approaches consensus \cite{vakrre03,caegsa06,daga08}. The consensus time was found to be a slowly increasing function of the number of states, saturating when the number of states tends to infinity \cite{stbapa12,pili16}. The geometry of the evolution to consensus depends on, amongst other things, the effective dimension of the interaction network. In two dimensions for example the logarithmic coarsening of the two-state VM can turn into algebraic ordering with an effective surface tension \cite{caegsa06,daga08}. 

Multi-state extensions of the NVM have been considered for example in \cite{francisco,ko18,valoba19}. In \cite{francisco} it was shown that there is no unique transition point in the multi-state NVM with more than two states $M>2$. Instead, the marginals of the stationary distribution change shape at their left and right edges at different threshold values of the noise strength. This is a consequence of a breaking of symmetry. In the two-state model, one has $P^{\rm st}(n_1)=P^{\rm st}(N-n_1)$ by constructions. In multi-state models, this symmetry no longer holds for the marginal distributions, $P_i^{\rm st}(n_i)$ for the number $n_i$ of individuals holding a particular opinion $i=1,\dots,M$. Instead, the marginal distributions for the $n_i$ are often concentrated on small but non-zero values.

As a separate aspect, the study of zealots in VMs has attracted attention. Zealots are agents who never change opinion. Their influence on `free individuals' (individuals who can change opinion) has been analyzed in variants of the VM \cite{mo03,mopere07,fues14,chepbaag15}, and in related models \cite{mo15,memozi16,memozi17,misz18,brrest18,brrest19,jesz20}. The presence of zealots gives rise to a broader phenomenology. In particular, in \cite{nagi} it was shown that zealots in the two-state NVM affect the nature of the transition between unimodal and bimodal states. Changes of shape of the stationary distribution can happen near $n_i=0$ and $n_i=N$ respectively at different ratios of noise and imitation strengths. As a necessary condition to this, the number of zealots for each of the two opinion states must not be the same \cite{nagi}. There is then again a breaking of symmetry in the stationary distribution. The transition is removed entirely when there is an equal number of zealots for each opinion state, and if these zealots affect all free individuals.  

The objective of this work is to study the effect of zealots in multi-state NVM. More specifically, we consider the NVM with all-to-all interaction (`mean-field'), and two types of individuals: free voters, who can be in any of the opinion states, and zealots. Our main aim is to understand how the combination of multiple states, noise and zealots affects the nature of the transition between multi-modal and unimodal states.

The remainder of the paper is set out as follows. Section \ref{model} contains the definitions of the model. In Section \ref{sec:K_eq_1} we analyse the baseline case in which all zealots affect the entire population of free agents. Our results are exact, and allow us to construct the phase diagram for the shapes of the marginals of the stationary distribution. In Section \ref{subpopulation} we then consider a more general situation in which free agents are dived into different communities, and where zealots in any one community only affect free agents in that community. This is shown to extend the range of possible phases. Theoretical predictions of the previous sections are compared against numerical simulations in Section \ref{simulations}. We summarise and discuss our work in Section \ref{summary}.

\section{Model definitions}
\label{model}

We consider population of $N$ `free' individuals and $Z$ zealots. At any one time, each individual holds one of $M$ opinions. We label opinion states $i=1,\dots,M$. Free individuals can change opinion in an imitation process, to be described below. The set of agents is divided into $K$ communities, which we label $k=1,\dots, K$. The community any one agent belongs to is fixed in time. We write $N\kk$ for the number of free agents in community $k$, and $Z\kk$ for the number of zealots in community $k$. The letters `Z' and `z' are pronounced {\em zet} throughout this paper, not {\em zee}. We have $N=\sum_{k=1}^K N\kk$, and $Z=\sum_{k=1}^K Z\kk$. We always assume $K\geq 1$, $M\geq 2$ and $N\geq M$

Each zealot is of a particular opinion state. We write $z_i\kk$ for the number of zealots of opinion $i$ in community $k$. Hence we have $Z\kk=\sum_{i=1}^M z_i\kk$. The number of zealots is fixed at the beginning and does not change with time.

We denote the number of free agents in community $k$ who are in opinion state $i$ by $n_i\kk$. We will use the words `opinion state' and `type' interchangeably, and refer to an individual who is of opinion $i$ as an agent of type $i$. The population of free agents has no further structure beyond the division into communities. The state of the system at any one time is therefore fully specified by the vectors $\bn\kk=(n_1\kk,\dots,n_M\kk)$, $k=1,\dots,K$. The total number of agents in opinion state $i$ is $n_i=\sum_{k=1}^Kn_i\kk$, and we have $N\kk=\sum_{i=1}^M n_i\kk$. We write $\mathbf{\underline n}=(\mathbf n^{(1)},\dots,\mathbf n^{(K)})$.

Changes of opinion occur following the usual rules of the multi-state noisy voter model. Free agents can change opinion by interacting with another individual (free voters or zealots) in the population. In this process a free agent copies the opinion state of the interaction partner. We assume that any free agent can interact with any other free agent in the population, regardless of the communities they belong to. Free agents in one community can however only be influenced by zealots in that community.

Spontaneous opinion changes of free voter are also possible. The rate by which such state changes occur is proportional to the model parameter $\varepsilon$. This quantity can be seen as a `mutation rate' \cite{francisco}. We assume that there is no particular ordering of the opinion states; when a mutation occurs the new opinion state of the agent is chosen at random and with equal probability from the applicable $M-1$ opinion states (the state the agent is in just prior to mutation is excluded). 

We assume that the dynamics unfolds in continuous time. The processes in the population can then be defined by the rates
\begin{equation}
  \label{eq:rates}
  T\kk_{i\to j}(\mathbf{\underline n})=r\frac{n_i\kk (n_j+z_j\kk)}{N+Z\kk}+\varepsilon n_i\kk.
\end{equation}
This quantity is the rate with which free agents of type $i$ in community $k$ convert into agents of type $j$. The first term on the right-hand side describes events in which an individual of type $i$ in community $k$ interacts either with a free agent of type $j$ of any community, or with a zealot of opinion $j$ influencing community $k$. As a consequence of this interaction the focal agent changes state to $j$. The coefficient $r>0$ in Eq.~(\ref{eq:rates}) describes the `imitation rate'. Its role is mainly to set the units of the time axis, and we will set $r=1$ throughout. We can then also think of the imitation dynamics as follows. An agent is chosen at random from the entire population. Suppose this individual is in community $k$. The individual then interacts with a partner chosen at random from the $N+Z\kk$ members in community $k$, and adopts that individual's opinion.

The second term in Eq.~(\ref{eq:rates}) describes spontaneous opinion changes in community $k$ from state $i$ to state $j$. These are taken to occur with per capita rate $\varepsilon$. 

The focus of our investigation will be on the total number of free agents holding any particular opinion. This is described by the vector $\mathbf n=(n_1,\dots,n_M)$. We note that only $M-1$ of the entries of $\mathbf n$ are independent, since $\sum_{i=1}^M n_i=N$. To describe the dynamics of these variables we introduce rates $T_i^+$ and $T_i^-$. These are the rates for events in which the total number of agents of type $i$ across communities is increased or reduced by one, respectively, $n_i\to n_i\pm 1$. The $T_i^\pm(\mathbf n)$ are obtained as
\begin{eqnarray}
  \label{eq:brates}
  T_i^+(\mathbf{\underline n})=&&\sum_{k=1}^K\sum_{j\ne i}T_{j\to i}\kk(\mathbf{\underline n}),
  \nonumber \\
  T_i^-(\mathbf{\underline n})=&&\sum_{k=1}^K\sum_{j\ne i}T_{i\to j}\kk(\mathbf{\underline n}).
\end{eqnarray}
Assuming a given configuration of zealots, and noting that $\sum_{j\neq i} n_j = N-n_i$ and Eq.~(\ref{eq:rates}), the rates $T_i^\pm$ for a fixed $i$ can be written in terms of the $n_i\kk$. That is to say knowledge of the $n_j\kk$, $j\neq i$ is not required to compute $T_i^+$ and $T_i^-$. The reason for this reduction is as follows: From the point of view of an individual of type $i$, other free individuals in the population are either also of type $i$, or in different state $j\neq i$. Imitation and mutation rates are uniform across types, and it is irrelevant for the birth-death rates for type $i$ how many of the other free agents belong to what types $j\neq i$. All that matters for the purposes of the imitation of other free agents is how many individuals are not of type $i$. If there are multiple communities, and if the configuration of zealots varies in the different communities then a breakdown of $n_i$ into the $n_i^{(k)}$ is required to formulate the transition rates $T_i^\pm$.

The imitation and mutation rates $r=1$ and $\varepsilon$ in our model do not vary across opinion states. In addition, we assume all-to-all interaction between free agents. The only minimal structure we allow is through the division of the population of free agents into communities, each influenced only by a subset of zealots. While many extensions are possible, we deliberately choose a relatively stylised setup in order to be able to systematically investigate the effects of the combination of noisy, multiple states and zealots in voter models.

In Sec.~\ref{sec:K_eq_1} we focus on the case in which there is only one single community ($K=1$). The phase diagram can then be obtained exactly. Subsequently, we will study a population consisting of $K=2$ communities (Sec.~\ref{subpopulation}). Approximations are then required to carry out the analysis. We test analytical predictions for both cases against numerical simulations in Sec.~\ref{simulations}.
 
\section{One single community: zealots affecting the entire population}
\label{sec:K_eq_1}
\subsection{Effective transition rates and marginals of the stationary distribution}
We focus on the case of one single community $K=1$. This means that zealots affect the entire population of agents. We have $n_i=n_i^{(1)}$ for all $i$, and $z_i=z_i^{(1)}$. The birth and death rates for individuals of type $i$ are obtained from Eqs.~\eqref{eq:rates} and (\ref{eq:brates}) as
\begin{eqnarray}
T_i^+(n_i)&=&\frac{(N-n_i)(n_i+z_i)}{N+Z}+\varepsilon(N-n_i), \nonumber \\
  \label{eq:bdratesall}
T_i^-(n_i)&=&n_i \frac{N-n_i+Z-z_i}{N+Z}+(M-1)\varepsilon n_i.
\end{eqnarray}
No approximation has been made to arrive at these expressions. The object $T_i^+(n_i)$ is the rate with which individuals of type $i$ are generated ($n_i\to n_i+1$), and the second, $T_i^-(n_i)$, is the rate with which individuals of opinion $i$ change to any other opinion ($n_i\to n_i-1$). The rates $T_i^\pm$ only depend on $n_i$, but not on the $n_j$ with $j\neq i$. This was observed  in \cite{francisco} in absence of zealots, and continues to be the case if zealots are present. We stress that this approach does not allow us to capture correlations between $n_i$ and $n_j$ for $i\neq j$. That is to say, from Eqs.~(\ref{eq:bdratesall}) [or Eqs. (\ref{eq:brates}) in more general] we cannot derive information about the joint statistics of the $n_i$, $i=1,\dots,M$. Instead our focus is on the shape of the marginal distributions for individual variables $n_i$.

Using well-known results for one-step processes \cite{ka92}, the marginal stationary distribution for $n_i$ is
\be\label{eq:statmarg}
P^{\rm st}_i(n_i)=\frac{\prod_{k=1}^{n_i} \frac{T_i^+(k-1)}{T_i^-(k)}}{1+\sum_{k=1}^N \prod_{\ell=1}^k \frac{T_i^+(\ell-1)}{T_i^-(\ell)}},
\ee
where $n_i=0,\dots,N$. As in \cite{francisco,nagi} the shape of these marginals is a good indicator of the overall stationary distribution of the population, and determines the different phases of the system. 

In order to characterise the shape of the marginals we first formulate the following lemma (a proof can be found in Appendix \ref{appen:1}): \\

\emph{Lemma 1}: The marginal stationary distribution $P^{\rm st}_i(n_i)$ in Eq.~(\ref{eq:statmarg}) has at most one extremum in $n_i\in\{1,\dots,N-1\}$. 
\\

As a consequence of this lemma, we can determine the qualitative shape of the marginals once we know whether an interior extremum is present or not, and if it is, whether this is a minimum or maximum. In order to do this, in turn, we only need to look at the behaviour of the function $P^{\rm st}_i$ near $n_i=0$ (`left edge') and near $n_i=N$ (`right edge'). We will also examine the central region of the distribution ($n_i$ near $n_i=N/2$).

\subsection{Right edge}

We start by looking at the right edge. Specifically, we would like to decide when $P_i^{\rm st}(n_i=N)$ is smaller or larger than $P^{\rm st}_i(n_i=N-1)$ respectively. This determines the `slope' of the distribution at the right edge.

In the stationary state, there is no net flux of probability between states $n_i=N-1$ and $n_i=N$, i.e., $P_i^{\rm st}(N-1)T_i^+(N-1)=P_i^{\rm st}(N)T_i^-(N)$. As a consequence, $P_i^{\rm st}(N-1)=P_i^{\rm st}(N)$ if and only if $T_i^+(N-1)=T_i^-(N)$. From this, and using Eqs.~(\ref{eq:bdratesall}) we find that the marginal distribution for $n_i$ changes sign at the right edge when
\begin{eqnarray}\label{eq:help_r}
&&\frac{1}{N+Z}\left[z_i(N+1)-N(Z-1)-1\right] \nonumber \\ && \qquad -\varepsilon\left[(M-1)N-1\right]=0.
\end{eqnarray}
This leads to a threshold value of $\varepsilon$
\be\label{eq:eps_r}
\varepsilon_{r,i} \equiv \frac{1}{N+Z}\frac{z_i-1- N(Z_{-i}-1)}{(M-1)N-1}
\ee
at which the slope of the distribution at the right edge changes sign. We have introduced the quantity $ Z_{-i}\equiv \sum_{j\neq i} z_j$. This the number of zealots of any type $j$, except $i$.

If the expression in Eq.~(\ref{eq:eps_r}) is positive then the shape of the marginal at the right edge changes as $\varepsilon$ crosses $\varepsilon_{r,i}$. More precisely, for $\varepsilon<\varepsilon_{r,i}$ the marginal distribution is an increasing function of $n_i$ near the right edge, and for $\varepsilon>\varepsilon_{r,i}$ it is decreasing. If $\varepsilon_{r,i}< 0$, then no change of shape can occur, one then always has a decreasing shape at the right edge, i.e., $P_i^{\rm st}(N)<P_i^{\rm st}(N-1)$.

We note that the square bracket multiplying $\varepsilon$ in Eq.~(\ref{eq:help_r}) is always positive (reflecting the fact that mutation acts in direction away from the edges of state space). This contribution can only be overcome by the imitation process if the term in the first square bracket is positive. This terms describes the net force due to imitation processes, and can be directed towards or away from the state $n_i=N$, depending on the number of zealots for the different opinion states. A change of shape can only occur if the net imitation force is towards opinion state $i$, and when it balances mutation.

If there are no zealots in the population at all ($z_i=0, Z_{-i}=0$) then $\varepsilon_{r,i}$ as defined in Eq.~(\ref{eq:eps_r}) is always positive, and the change of shape of the marginal at the right edge occurs when  $\varepsilon=\varepsilon_{r,i}=\frac{1}{N}\frac{N-1}{(M-1)N-1}$. 

However if $Z_{-i}>0$, i.e. if there are zealots of a any type $j\neq i$ then an ${\cal O}(N)$ number of zealots for opinion $i$ is required to generate a net imitation force towards state $i$. A change of shape can then occur when $\varepsilon=\varepsilon_{r,i}$. If there are not sufficiently many zealots of type $i$, then mutation away from state $i$ combined with the zealots for opinions $j\neq i$ dominate, and $P_i^{\rm st}(N-1)>P_i^{\rm st}(N)$.

We now focus on the case with equally many zealots for the different opinions states ($z_i\equiv Z/M$ for all $i$). The condition for the existence of a transition, $\varepsilon_{r,i}>0$, becomes
\be
z_i<\frac{N-1}{N(M-1)-1}.
\ee
The right-hand side is evidently strictly smaller than one for $M\geq 3$. This means that the transition at the right edge cannot occur in noisy voter models with three or more opinion states and a non-zero number of zealots equally distributed across the opinion states. Similar behaviour was previously noted for $M=2$ in \cite{nagi}. Our analysis shows that this result holds for a general number of opinion states.

\subsection{Left edge}
The shape of the distribution near the left edge ($n_i=0$) is determined by the sign of the following quantity,
\BE
\nonumber
T_i^+(0)-T_i^-(1)= && \frac{1}{N+Z}\left[Nz_i - N +1 -Z+z_i\right]\\ &&+\varepsilon\left[N-M+1\right].
\EE
The second term on the right-hand side is always positive (mutation is directed towards the centre of state space). As a consequence, a shape-change as a function of $\varepsilon$ can only occur if $Nz_i-N+1-Z_{-i}<0$. For a given value of $Z_{-i}$, this means that there must not be too many zealots of type $i$. A change of shape at the left edge then occurs at $\varepsilon=\varepsilon_{\ell,i}$, with
\be
\label{eq:eps_l}
\varepsilon_{\ell,i}\equiv \frac{1}{N+Z}\frac{Z_{-i}-1-N(z_i-1)}{N+1-M}.
\ee
If the expression on the right-hand side is negative, then the marginal distribution is always increasing at the left edge [$P_i^{\rm st}(1)>P_i^{\rm st}(0)$]. 

In the case of balanced numbers of zealots, $z_i=Z/M$ for all $i$, the condition $\varepsilon_{i,\ell}>0$ can be written as $\frac{Z}{M}(N+1)<N-1+Z$. For $M=2$ this turns into $Z/2<1$, so the shape-change is possible only if $Z=0$, see \cite{nagi}. For general $M$, we require
\be
\frac{Z}{M}<\frac{N-1}{N+1-M}.
\ee
We note that this inequality turns into $Z<M(M-1)$ for $M=N$. This means that a change of shape at the left edge can occur for quite a large number of zealots.

\subsection{Central region and symmetry of the marginal distribution}

We now look at the shape of the marginal distribution for $n_i$ in the central region near $n_i=N/2$. In particular we determine the conditions for which any possible extremum of $P_i^{\rm st}(n_i)$ is at $N/2$. To do this, we first note the following:

\medskip

\emph{Lemma 2:} The marginal distribution $P_i^{\rm st}$ is found to be symmetric [$P_i^{\rm st}(n_i)=P_i^{\rm st}(N-n_i)$ for all $n_i\in\{0,1\dots,N\}$] for $M>2$ when $\varepsilon$ takes the value
  \begin{equation}\label{eq:eps_c}
  \varepsilon_{c,i}\equiv \frac{z_i-Z_{-i}}{(M-2)(N+Z)}.
\end{equation}

\medskip

This can be seen by direct algebra as follows. The condition in Eq.~(\ref{eq:eps_c}) implies $T_i^+(n_i)=T_i^-(N-n_i)$ for the rates in Eq.~(\ref{eq:bdratesall}). The symmetry of the stationary distribution then follows. 

We have already shown that the marginal distribution for $n_i$ can have at most one extremum in the interiour (Lemma 1). Therefore, at $\varepsilon=\varepsilon_{c,i}$ the distribution must either be flat, or have an extremum at $n_i=N/2$.

Conversely we can show for even $N$ that the distribution $P_i^{\rm st}$ takes its extremal value at $n_i=N/2$ in a corridor of values for $\varepsilon$ around $\varepsilon_{c,i}$. The corridor has a width of order ${\cal O}(N^{-1})$. The existence of such a corridor is a consequence of the discreteness of the variable $n_i$. Further details can be found in Appendix \ref{sec:proof2}.

For $M=2$ the distribution $P_1(n_1)$ is symmetric only when $z_1=z_2$ \cite{nagi}. If there are more than two opinion states $M>2$, and noting $Z_{-i}=Z-z_i$, the expression for $\varepsilon_{c,i}$ in Eq.~\eqref{eq:eps_c} is positive only when $z_i>Z/2$. If $\varepsilon_{c,i}$ is negative then the marginal for $n_i$ does not have an extremum at the centre for any choice of $\varepsilon$ \cite{note1}. A zero or negative value for $\varepsilon_{c,i}$ is for example found if there are equally many zealots for each opinion state, $z_i=Z/M$ and $M\ge 2$.

We also note that there can be at most one possible opinion $i$ for which $z_i>Z/2$. As a consequence, only at most one of the numbers $\varepsilon_{c,i}$, $i=1,\dots,M$ can be positive. This must then be the opinion state with the most number of zealots. Only the marginal distribution for this opinion state can have a maximum or minimum at the centre.

\subsection{Phase diagram: shapes of the marginals}\label{sec:pg}
\subsubsection{General structure of the possible phases}
\begin{figure}[t!]
  \centering
  \includegraphics[width=.45\textwidth]{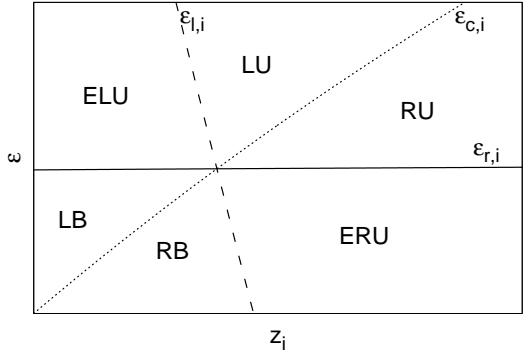}
  \caption{Illustration of the possible phases for the behaviour of the marginal stationary distribution for opinion state $i$. This is for the model with a single community, $k=1$. We show the phase diagram in the $z_i\--\varepsilon$ plane, assuming that $Z_{-i}$ remains fixed. Not all phases are necessarily accessible with physically meaningful parameters (see text). The solid line shows $\varepsilon_{r,i}$, the dashed line is $\varepsilon_{\ell,i}$, and the dotted line represents $\varepsilon_{c,i}$. The lines divide parameter space into six regions, with different shapes of the  marginal probability distribution of opinion $i$: LU left unimodal, ELB  extreme left bimodal, LB left bimodal, RB right bimodal, ERU extreme right unimodal, and RU right unimodal. Further details can be found in the text.}
  \label{figt:01}
\end{figure}
For fixed $N, M$ and $Z_{-i}$, the phase diagram for the marginal distribution $P_i^{\rm st}$ can be illustrated using the lines for $\varepsilon_{r,i}, \varepsilon_{\ell,i}$ and $\varepsilon_{c,i}$ in the $z_i\--\varepsilon_i$ plane. These lines are given by the expressions in Eqs.~(\ref{eq:eps_r}), (\ref{eq:eps_l}) and (\ref{eq:eps_c}) respectively. 

Direct algebra shows that the three lines intersect at the point given by  
\begin{eqnarray}
  \label{eq:z_eps_star}
  z_i^*=&&\frac{Z_{-i}+M-2}{M-1}, \nonumber \\ \varepsilon^*=&&\frac{1-Z_{-i}}{(M-1)(N+1)+MZ_{-i}-1},
\end{eqnarray}
and that there are no other intersection points between any of the lines.  

As an aside, it is interesting to note the following (a proof can be found in Appendix \ref{appen:1}):

\medskip

\emph{Lemma 3}: At the intersection point in Eq.~(\ref{eq:z_eps_star}) the marginal stationary distribution function for $n_i$ is flat. 

\medskip

We illustrate the general topology of the resulting phase diagram in Fig.~\ref{figt:01}. We note that not all phases can be physically realised for all choices of $N, M$ and $Z_{-i}$. This includes situations in which the values for $\varepsilon_{r,i}, \varepsilon_{\ell, i}$ or $\varepsilon_{c,i}$ are negative, such that the corrresponding phase lines are in an unphysical part of the phase diagram. We also note that $z_i$ can only take integer values. In the diagram in Fig.~\ref{figt:01} we ignore this for the time being and treat $z_i$ as continuous. We discuss the limitations due to the restriction to integers below.

To understand the diagram we note that the marginal for opinion $i$ has a maximum at $n_i=0$ for $\varepsilon<\varepsilon_{\ell,i}$, otherwise it is an increasing function at the left edge. It has a maximum at $n_i=N$ for $\varepsilon<\varepsilon_{r,i}$, and is a decreasing function at the right edge otherwise. In the region to the upper right of the diagram, where $\varepsilon>\varepsilon_{r,i}$ and $\varepsilon>\varepsilon_{\ell,i}$ the marginal is therefore increasing near $n_i=0$ and decreasing near $n_i=N$. As a consequence, it has a maximum in the interior $n_i=1,\dots,N-1$. This is the combined region of the phases marked LU and RU, where U indicates a unimodal shape. The region is divided into LU and RU by the line $\varepsilon=\varepsilon_{c,i}$. Along this line the distribution is symmetric, and the interior maximum is at $n_i=N/2$. If $\varepsilon>\varepsilon_{c,i}$ then the maximum is found at $n_i<N/2$ (LU phase, `left unimodal'), for $\varepsilon<\varepsilon_{c,i}$ the maximum is at a value $n_i>N/2$ (RU phase, 'right unimodal').

Conversely, in the regions labelled LB and RB in the lower left region of the diagram, $\varepsilon<\varepsilon_{r,i}$ and $\varepsilon<\varepsilon_{\ell,i}$. The marginal distribution is then bimodal as indicated by the letter `B', it has a minimum in the interior, and maxima at $n_i=0$ and $n_i=N$. We note that Lemma 1 only forbids multiple extrema in the interior ($n=1,\dots,N-1$), but not at the edges $n_i=0$ and $n_i=N$. The line $\varepsilon=\varepsilon_{c,i}$ divides the region into the LB and RB phases. Along the line the distribution is symmetric, the minimum is at $n_i=N/2$, and the maxima at $n_i=0$ and $n_i=N$ have equal height, $P_i^{\rm st}(n_i=0)=P_i^{\rm st}(n_i=N)$. In the LB phase the distribution takes a higher value at $n_i=0$ than at $n_i=N$ (`left bimodal'), and in the RB phase the higher maximum is at $n_i=N$ (`right bimodal').

The remaining regions in the phase diagram are those with $\varepsilon>\varepsilon_{r,i}, \varepsilon<\varepsilon_{\ell,i}$ and vice versa. In the diagram in Fig.~\ref{figt:01} these are the regions in the upper left and lower right respectively. These are marked ELU and ERU (`extreme left unimodal', `extreme right unimodal'). In the ELU phase the marginal distribution for $n_i$ is decreasing across the entire range $n_i=0,\dots, N$, and has a single maximum (`unimodal') at the extreme left ($n_i=0$). In the ERU phase the distribution is increasing throughout and has a single maximum at the extreme right ($n_i=N$).

Physically, the phase diagram can be understood as follows. Suppose we fix $z_i$ and $Z_{-i}$, along with $N$ and $M$. If $\varepsilon$ is sufficiently large (larger than $\varepsilon_{\ell,i}, \varepsilon_{c,i}$ and $\varepsilon_{r,i}$ for this particular value of $z_i$), then the system is dominated by the noise component of the dynamics (the mutation term). In the extreme case $\varepsilon\to\infty$, imitation plays no role at all, and all opinions are equally represented on average. All marginals are unimodal, and $n_i=N/M$ on average for all $i$. The maximum of the unimodal distribution is near this value $n_i=N/M$, and the marginals are  therefore of the LU shape.

Suppose now we are in the LU phase and have $z_i>0$. We now move counter-clockwise in the phase diagram in Fig.~\ref{figt:01}. If we reduce the number of zealots $z_i$ influencing agents of type $i$, but keep the rest of the parameters fixed (including $Z_{-i}$), then the maximum of the distribution for $n_i$ moves to the left towards $n_i=0$. When  $\varepsilon=\varepsilon_{\ell,i}$ is reached,  the system enters the extreme left unimodal (ELU) phase. The marginal is a decreasing function of $n_i$, with its maximum at $n_i=0$. 

We now move downward in the phase diagram by decreasing $\varepsilon$. This tends to shift probability towards the edges $n_i=0$ and $n_i=N$. When $\varepsilon$ crosses $\varepsilon_{r,i}$, the distribution changes slope (from decreasing to increasing) at the right edge, and the system enters the LB phase. The marginal for $n_i$ is now bimodal, with maxima at $n_i=0$ and $n_i=N$, and a minimum at a value $n_i>N/2$. 

Next, we move to the right in the phase diagram by adding further zealots of type $i$ (while keeping $Z_{-i}$ constant). The minimum of the distribution then moves towards smaller values of $n_i$, and reaches $n_i=N/2$, when $\varepsilon=\varepsilon_{c,i}$. Beyond this value, the marginal takes its minimum at values $n_i<N/2$, and the system is in the right bimodal (RB) phase. We have $P_i^{\rm st}(n_i=N)>P_i^{\rm st}(n_i=0)$

Further increasing $z_i$ raises the distribution at larger values $n_i\approx N$, and lowers it near $n_i=0$. When $\varepsilon=\varepsilon_{\ell,i}$, the marginal becomes an increasing function of $n_i$ throughout with its maximum at $n_i=N$, and the system is in the extreme right unimodal phase (ERU). 

Increasing the noise strength $\varepsilon$, the maximum is shifted away from $n_i=N$. The system is in the right unimodal (RU) phase. The marginal is a unimodal function with its maximum at a value $n_i>N/2$. 

Removing zealots of type $i$ finally, shifts the position of the maximum to smaller values of $n_i$. At $\varepsilon=\varepsilon_{c,i}$ the maximum is found at $n_i=N/2$ and upon further reduction of the noise strength the system enters the LU phase.

\subsubsection{Not all phases are always realised}\label{sec:unphysical}
Not all six phases shown in Fig.~\ref{figt:01} are physically feasible for all choices of $N, M$ and $Z_{-i}$. Phases can for example become unphysical when they require formally negative values of $\varepsilon$, or when there are no integer values for $z_i$ in the respective region in the phase diagram. 

We illustrate this for the RB phase. In order to be in this phase, $\varepsilon$ must be such that $\varepsilon<\varepsilon_{r,i}$, $\varepsilon<\varepsilon_{\ell,i}$ and $\varepsilon>\varepsilon_{c,i}$. We therefore require $\varepsilon_{r,i}>0$ and $\varepsilon_{\ell,i}>0$. Using Eqs.~(\ref{eq:eps_r}) and (\ref{eq:eps_l}) this means that $z_i-1>N(Z_{-i}-1)$ and $Z_{-i}-1>N(z_i-1)$. This is only possible simultaneously when $z_i=Z_{-i}=0$. If that is the case however, then $\varepsilon_{c,i}=0$ [Eq.~(\ref{eq:eps_c})]. We therefore conclude that the conditions to be strictly in the RB phase can never be fulfilled for $M>2$. For $M=2$, the RB phase is not present either as was shown in \cite{nagi}.

Further insight can be drawn from the limit in which the number of free agents is much larger than the number of opinions and the number of zealots, $N\gg M,Z$. Recalling $Z_{-i}=Z-z_i$, the expressions for $\varepsilon_{r,i}, \varepsilon_{\ell,i}$ and $\varepsilon_{r,i}$ then become linear in $z_i$
\begin{eqnarray}
  \label{eq:clra}
  && \varepsilon_{r,i}\approx \frac{1}{(M-1)N}(1-Z_{-i}), \nonumber \\
  && \varepsilon_{\ell,i}\approx \frac{1}{N}(1-z_i), \nonumber \\
  && \varepsilon_{c,i}\approx \frac{z_i-Z_{-i}}{(M-2)N}.
\end{eqnarray}
It is then manifest that several of the phases in Fig.~\ref{figt:01} cannot be realised if $z_{i}\ge 1$ or $Z_{-i}\ge 1$. This is a consequence of the global influence of the zealots on all free agents, an aspect that we relax in the following section.

\section{Zealots affecting only sub-populations}
\label{subpopulation}

We now generalise the setup to situations in which the population of agents divides into $K\geq 1$ communities. Free agents can interact across communities, but zealots in any one community can only influence free agents in that community.

\subsection{Approximation for effective dynamics and phase diagram}
For models with more than one community the birth-death rates, $T_i^\pm$, for type $i$ in Eq.~(\ref{eq:brates}) cannot be expressed only in terms of $n_i$. Instead number of free agents of type $i$ in each of the communities is needed, i.e., $T_i^\pm=T_i^\pm(n_i^{(1)},\dots, n_i^{(k)})$.  This complicates further the analysis, and we therefore characterise the shape of the resulting marginal distributions using an approximation similar to the one in \cite{nagi}. The approximation is justified retrospectively through comparison against simulations.

The fundamental assumption underpinning the approximation is that the fraction of free agents who are in a given opinion state $i$ is constant across communities, i.e. 
\begin{equation}
  \label{eq:approx}
  \frac{n_i\kk}{N^{(k)}} \approx \frac{n_i}{N},~\mbox{for all~}k. 
\end{equation}
Making this assumption, the rates in Eqs.~\eqref{eq:brates}  become 
\begin{eqnarray}
  \label{eq:bdrates_allk}
  T_i^+(n_i)&=& \frac{(N-n_i)(\alpha n_i+\tilde z_i)}{N+Z}+\varepsilon(N-n_i), \nonumber \\
  T_i^-(n_i)&=&n_i \frac{\alpha(N-n_i)+\tilde Z-\tilde z_i}{N+Z}+(M-1)\varepsilon n_i, \nonumber \\
\end{eqnarray}
where we have introduced
\be\label{eq:z_tilde}
       \tilde z_i=\sum_k q\kk     z_i^{(k)},
\ee
with coefficients 
\be\label{eq:qk}
q\kk\equiv \frac{N^{(k)}/(N+Z^{(k)})}{N/(N+Z)}.
\ee
We have also written $\tilde Z=\sum_i \tilde z_i$, and $\alpha \equiv \sum_k q\kk$ in Eqs.~(\ref{eq:bdrates_allk}).

The relations in Eqs.~(\ref{eq:z_tilde}) and (\ref{eq:qk}) have a particularly straightforward interpretation when $N\gg Z$. We then have $q\kk\approx N^{(k)}/N$, and therefore $\tilde z_i = \sum_k (N^{(k)}/N) z_i\kk$. We can therefore think of $\tilde z_i$ as an effective number of zealots for opinion $i$. It is a weighted average of the number of zealots of type $i$ across communities, where each community is weighted according the the number of free agents in the community. The pre-factor $\alpha$ reduces to unity for $N\gg Z$. For general values of $N$ and $Z$, the number of effective zealots for opinion $i$ continues to be a linear combination of the number of zealots in each community, but now with coefficients given in Eq.~(\ref{eq:qk}). The  approximation in Eq.~\eqref{eq:approx} therefore leads to a model with an effective number of zealots who influence the entire population of free agents. These zealots arise from distributing the zealots in the different communities across the entire population with suitable weights. As a consequence of this the effective number $\tilde z_i$ of zealots for any one opinion $i$ is not necessarily an integer number.

The rates in Eqs.~(\ref{eq:bdrates_allk})  reduce to those in Eqs.~(\ref{eq:bdratesall}) in the case of one single community $K=1$. For $K\geq 2$ they remain of a form which is very similar to those in Eqs.~\eqref{eq:bdratesall}.  The main differences are the coefficient $\alpha$ inside the imitation term, and the replacement of $z_i$ by $\tilde z_i$. 
\begin{figure*}[t!!]
  \centering 
  \includegraphics[width=.45\textwidth]{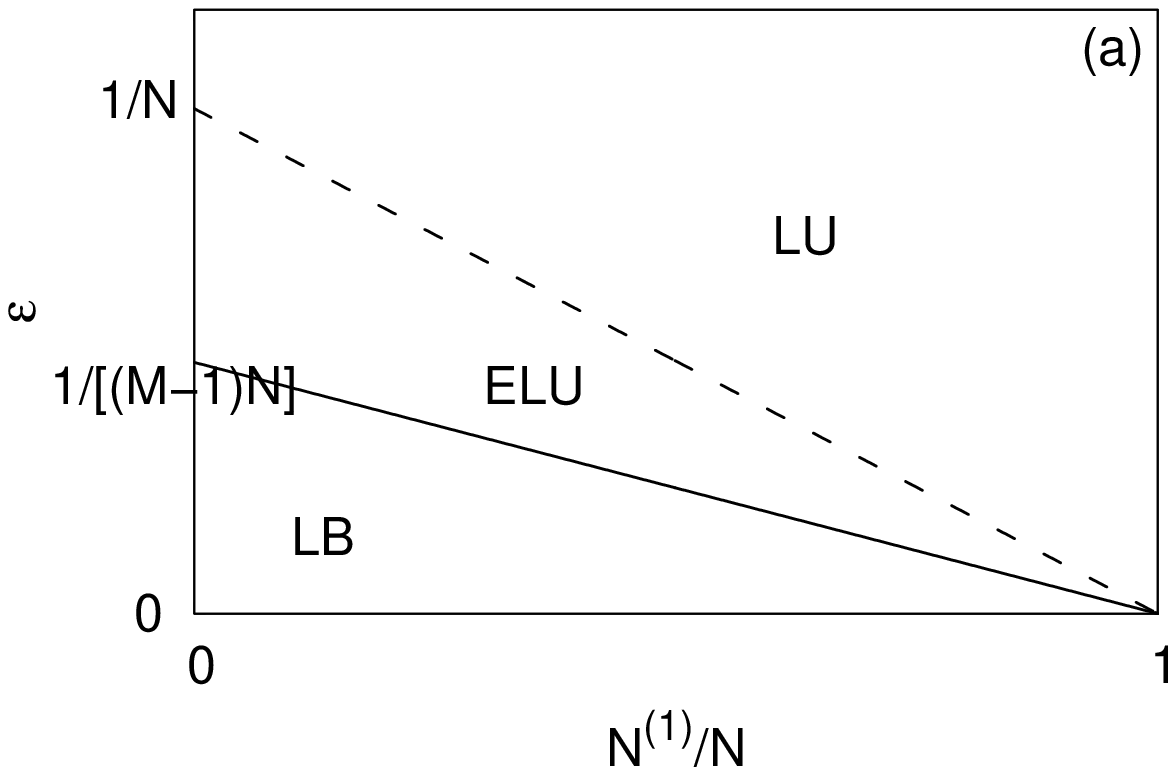}
  \includegraphics[width=.45\textwidth]{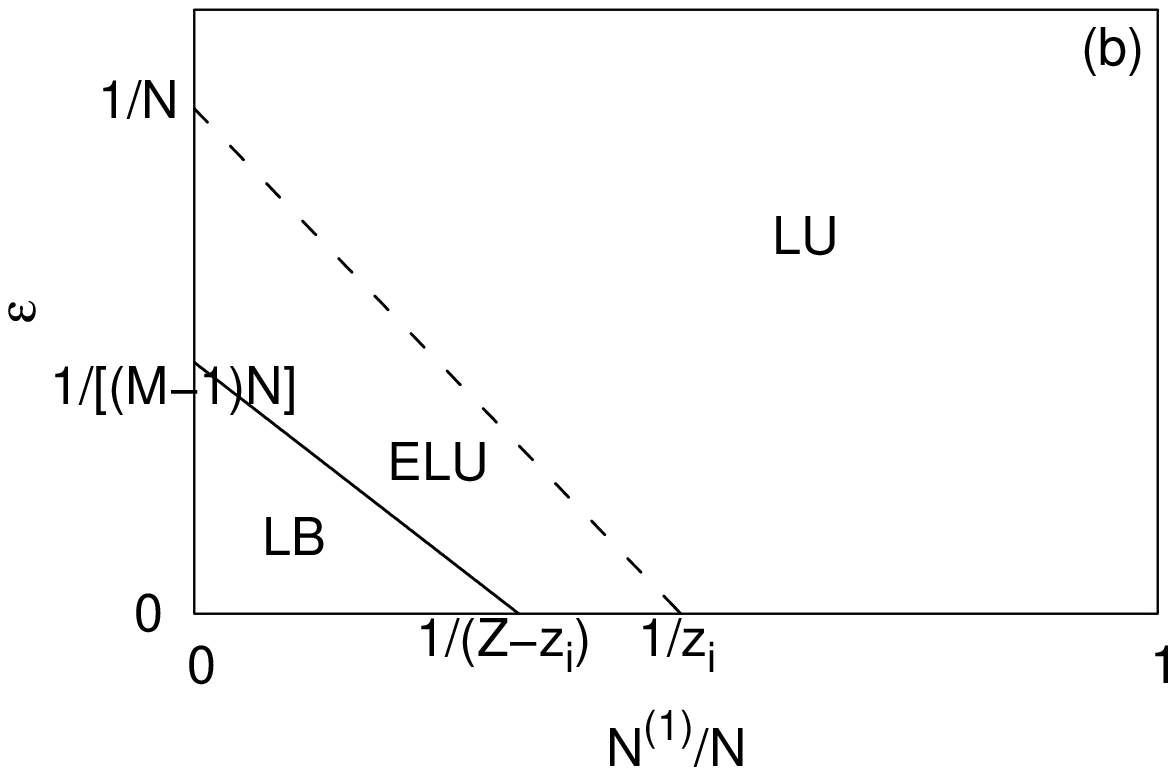}
  \caption{Examples of the phase diagram for the shape of the stationary marginal distribution $P_i^{\rm}(n_i)$ for the number of agents holding opinion $i$. This is for a model with two communities ($K=2$), and with zealots only in community $k=1$. The phase diagrams shown are for (a) $z_i=Z/2$ and (b) $z_i<Z/2$. Only the LU, ELU and LB phases can then be realised. The solid line in each panel is $\varepsilon_{r,i}$, the dashed one is $\varepsilon_{\ell,i}$, and the dotted line $\varepsilon_{c,i}$.
  \label{phasediagrampartinf1}}
\end{figure*}

As a consequence, the analysis proceeds along very similar lines as for the model with a single community in Section~\ref{sec:K_eq_1}, and the general structure of the phase diagram remains unchanged. The expressions for $\varepsilon_{r,i}, \varepsilon_{\ell,i}$ and $\varepsilon_{c,i}$ can be obtained from those in Eqs.~(\ref{eq:eps_r},\ref{eq:eps_l}) and (\ref{eq:eps_c}) by the replacement $N+Z\to (N+Z)/\alpha$ in the denominator of each expression, and $z_i\to \tilde z_i/\alpha, Z_{-i}\to \tilde Z_{-i}/\alpha$, where $\tilde Z_{-i}\equiv \tilde Z-\tilde z_i$. As an example, we discuss a model with two communities in the next section.

\subsection{Partial influence: Two communities ($K=2$)}

\begin{figure*}[t]
  \centering
  \includegraphics[width=.45\textwidth]{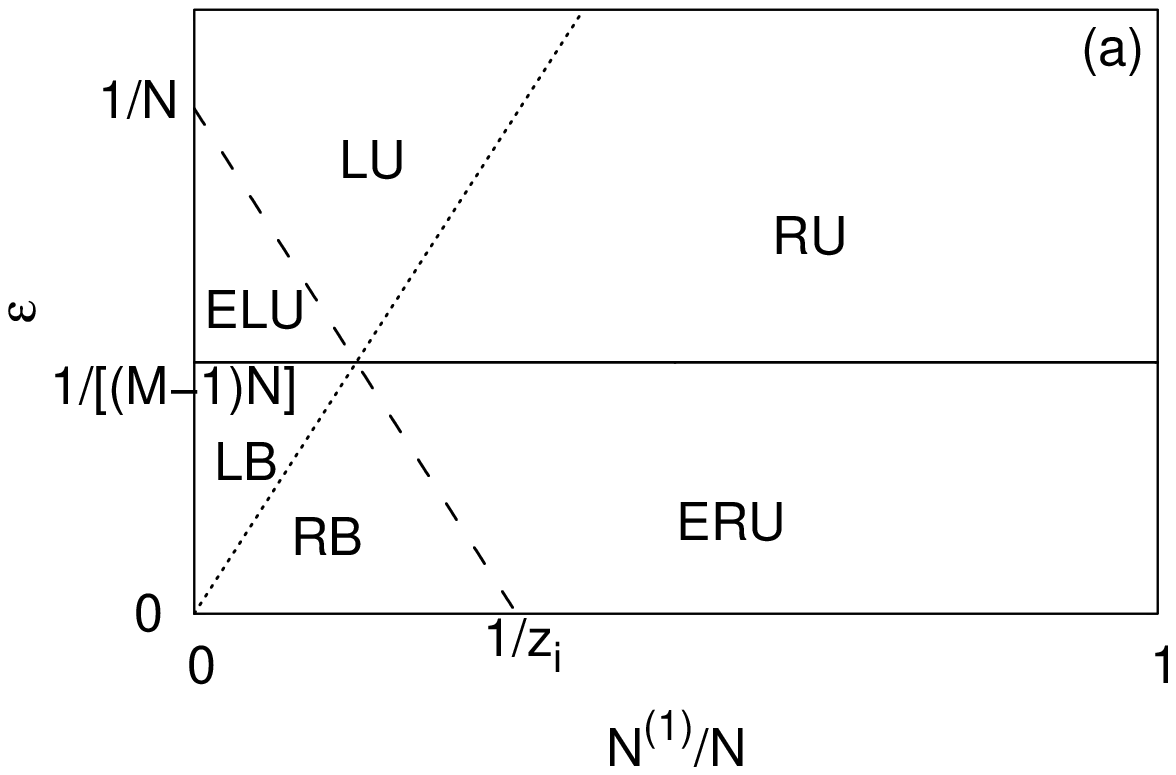}
  \includegraphics[width=.45\textwidth]{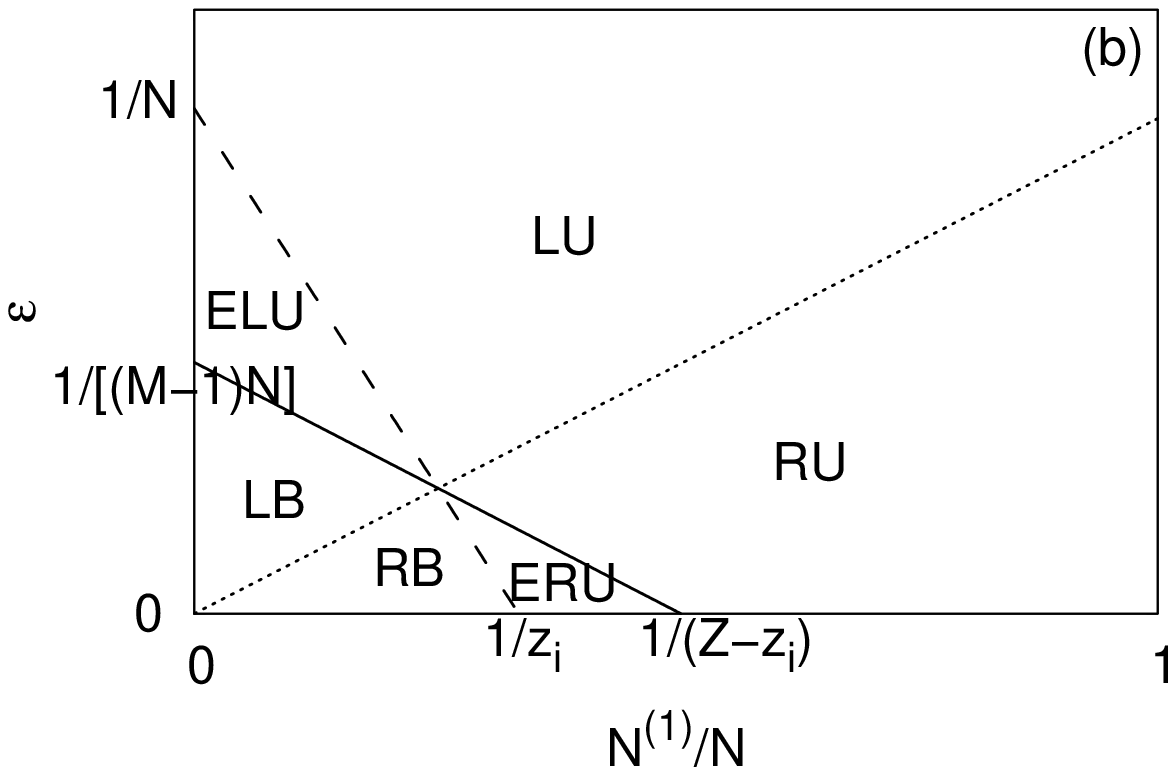}
  \caption{Examples of the phase diagram for the shape of $P_i^{\rm st}(n_i)$ for the model with two communities, and with zealots only in community $k=1$. In contrast with Fig.~\ref{phasediagrampartinf2} we now have $z_i> Z/2$, (a) $z_i=Z$ and (b) $Z/2<z_i<Z$. All six phases shown in Fig.~\ref{figt:01} can then be physically realised. The solid line in each panel is $\varepsilon_{r,i}$, the dashed line is $\varepsilon_{\ell,i}$, and the dotted line $\varepsilon_{c,i}$.\label{phasediagrampartinf2}}
 
\end{figure*}
\begin{figure*}[]
  \centering
  \includegraphics[width=.45\textwidth]{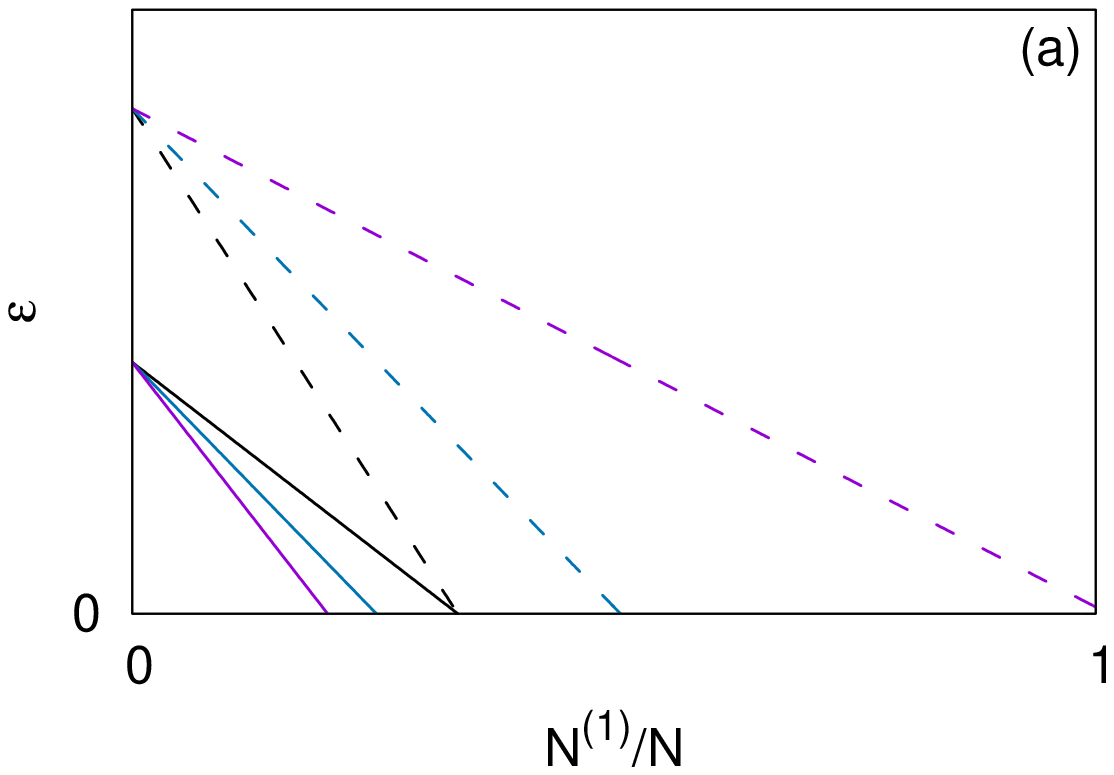}
  \includegraphics[width=.45\textwidth]{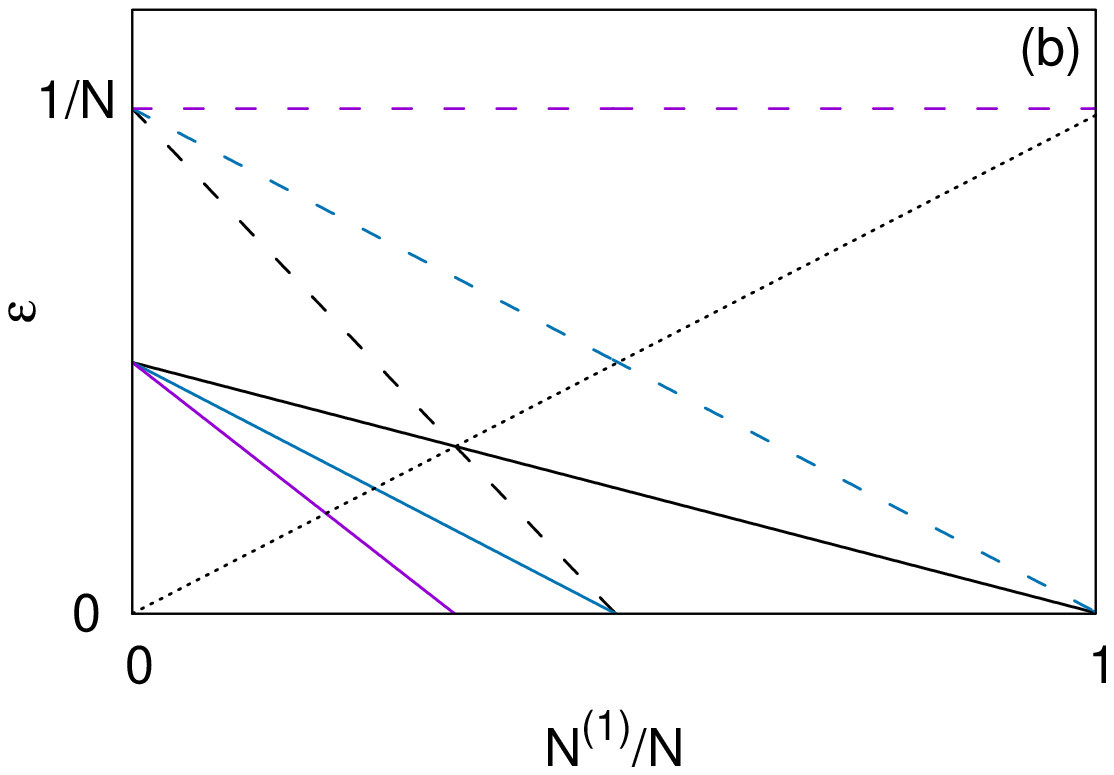}
  \caption{Representations of two topologically different types of phase diagram for a model with $K=2$ communities and $M=3$ opinion states. There are no zealots in community $k=2$. Numbers of zealots are such that $z_1^{(1)}>z_2^{(1)}>z_3^{(1)}$. The solid lines are $\varepsilon_{r,i}$ ($i=1,2,3$ from top to bottom), the dashed lines $\varepsilon_{\ell,i}$ ($i=1,2,3$ from bottom to top). The opinion state for each line is also indicated as color: black $i=1$, blue $i=2$, and purple $i=3$. The dotted line in panel (b) is $\varepsilon_{c,i}$ for the opinion with the most zealots ($i=1$). Panel (a): The number of zealots for opinion state $i=1$ is  $z_1^{(1)}\le Z/2$. Panel (b): $z_1^{(1)}>Z/2$. }
  \label{fig:-1}
\end{figure*}
We now look at a population consisting of two communities, $K=2$. Zealots are only present in the first community, but not in the second, i.e. $Z^{(2)}=0$. We then have $z_i=z_i^{(1)}$ and similarly, $Z=Z^{(1)}$ and $Z_{-i}=Z_{-i}^{(1)}$. We focus on the shape of the marginal distributions $P_i^{\rm st}(n_i)$, where $n_i$ is the total number of agents of type $i$ in both communities, $n_i=n_i^{(1)}+n_i^{(2)}$.

Similar to the one-community case, the marginal for opinion state $i$ is a symmetric function of $n_i$ [$P_i^{\rm st}(n_i)=P_i^{\rm st}(N-n_i)$ for $n_i=0,\dots, N$] when $\varepsilon=\varepsilon_{c,i}$. For fixed $N, M$ and $Z_{-i}$ the lines for $\varepsilon_{r,i}, \varepsilon_{\ell,i}$ and $\varepsilon_{c,i}$ intersect in one single point. If this occurs at physical parameters the marginal distribution for $n_i$ is flat. Further properties of the phase lines are discussed in Appendix \ref{appen:2}.

We find that there are two topologically distinct phase diagrams for the shape of the marginal for a particular opinion $i$. These are illustrated in the $N^{(1)}/N\--\varepsilon$ plane in Figs.~\ref{phasediagrampartinf1} and \ref{phasediagrampartinf2}.

Fig.~\ref{phasediagrampartinf1} shows cases for which  $z_i\le Z/2$. One then has $\varepsilon_{i,c}<0$, and only the left bimodal (LB), extreme left unimodal (ELU), and left unimodal (LU) phases are realised. 

In Fig.~\ref{phasediagrampartinf2} we have $z_i>Z/2$, and all six phases described in Sec.~\ref{sec:pg} can be realised. In particular, the intersection point of the lines $\varepsilon_{1,r},\varepsilon_{1,\ell}$ and $\varepsilon_{1,c}$ can be reached with physical control parameters.  An example of this will be shown in Sec.~\ref{sec:shape_marginals}.

The phase diagrams in Figs.~\ref{phasediagrampartinf1} and \ref{phasediagrampartinf2} indicate the shape of a single marginal for opinion state $i$. The situation becomes more complicated if we look at combinations of shapes for the marginals for different opinion states. In particular, if the number of zealots is different for the different opinion states, then the marginals for the different $n_i$ can have different shapes. Examples of this are shown in Fig.~\ref{fig:-1}, where we focus on a model with $M=3$ opinion states, $i=1,2,3$, and with decreasing numbers of zealots, $z_i=z_i^{(1)}$, from $i=1$ to $i=3$ ($z_1>z_2>z_3$). The diagrams show the phase lines $\varepsilon_{r,i}, \varepsilon_{\ell,i}$ and $\varepsilon_{c,i}$ in the $N^{(1)}/N \-- \varepsilon$ plane for $i=1,2,3$. A  number of combinations of shapes for the different marginals can then be found. The number of combinations is particularly high in the example shown in Fig.~\ref{fig:-1}(b).

\section{Numerical simulations}
\label{simulations}

We now compare the theoretical predictions for the phase diagrams against numerical simulations. We focus on the case of partial influence in Sec.~\ref{subpopulation}. Simulations are of the model defined by the rates in Eq.~\eqref{eq:rates}, and are carried out using the Gillespie algorithm \cite{gillespie}.
\subsection{Shape of marginals}\label{sec:shape_marginals}
\begin{figure}[t!]
  \centering
  \includegraphics[width=.235\textwidth]{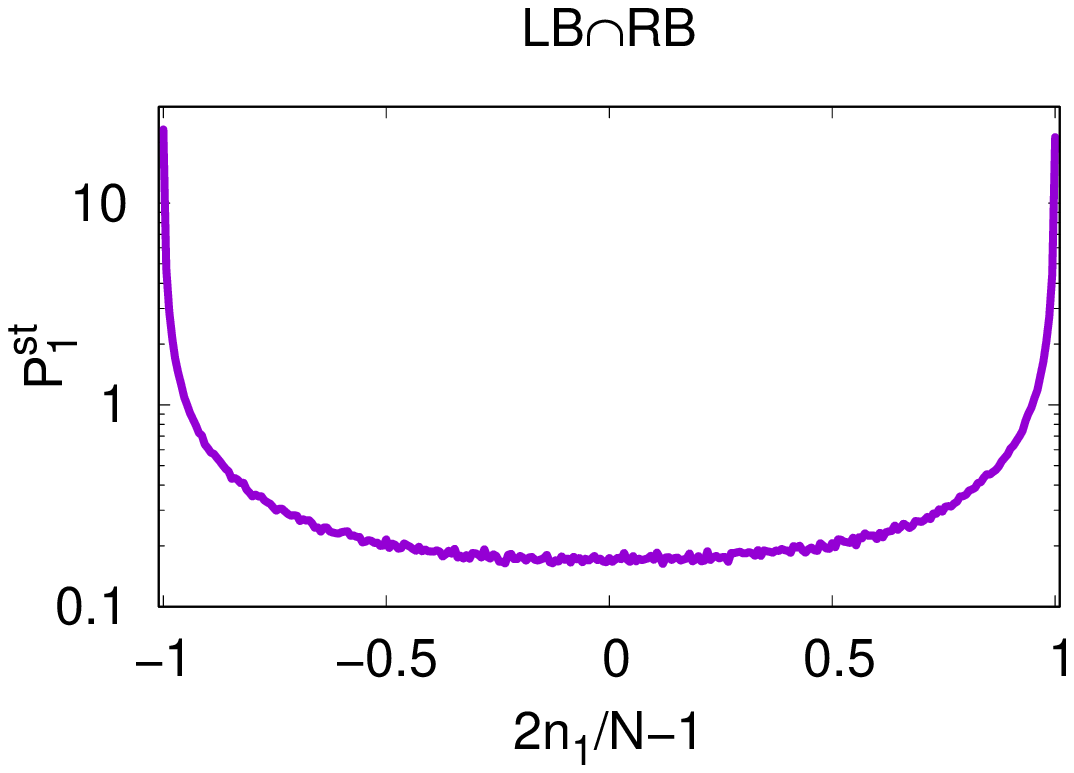}
  \includegraphics[width=.235\textwidth]{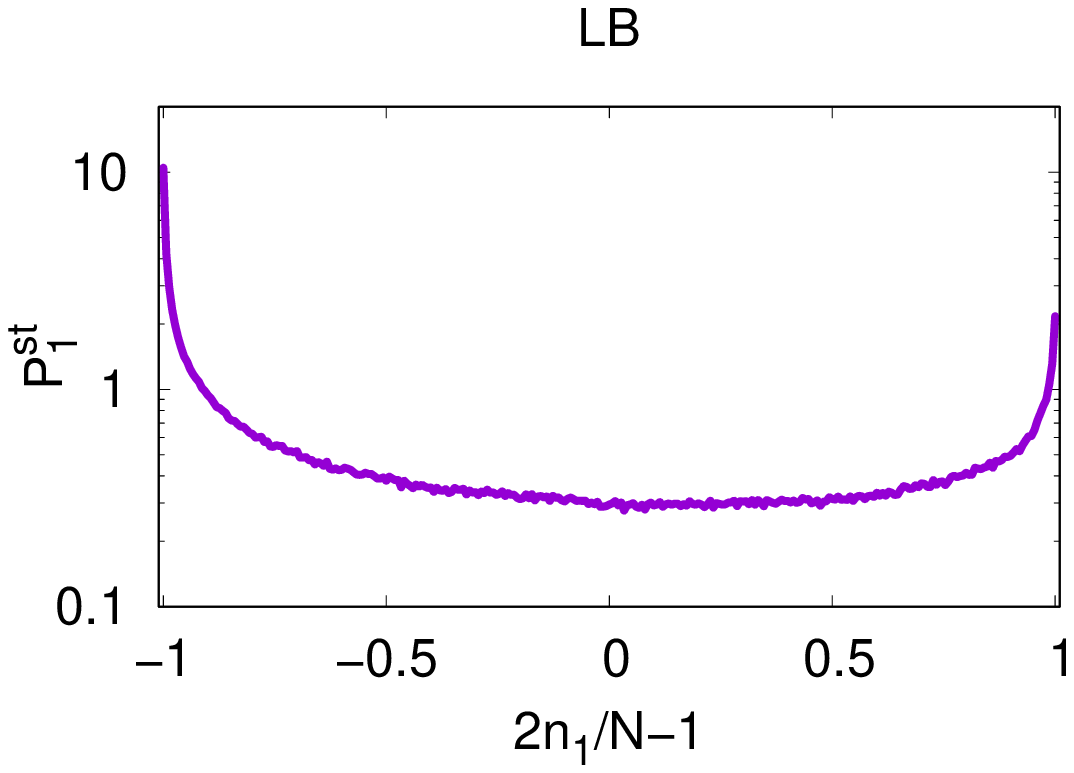}
  \includegraphics[width=.235\textwidth]{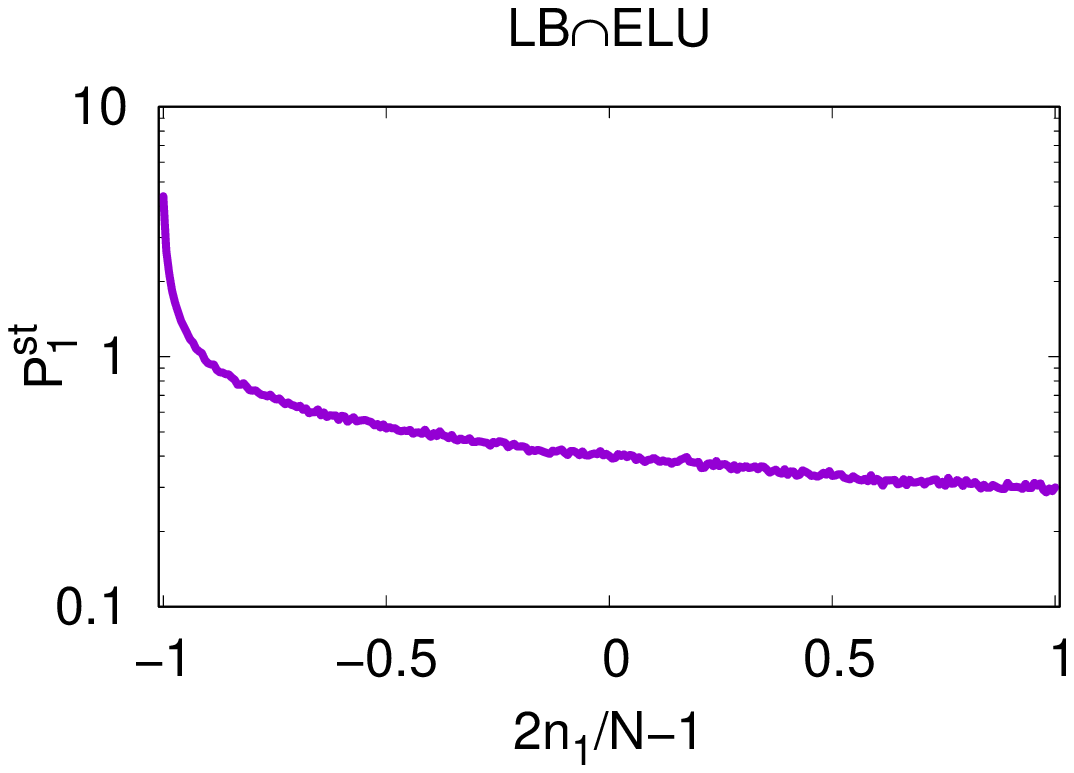}
  \includegraphics[width=.235\textwidth]{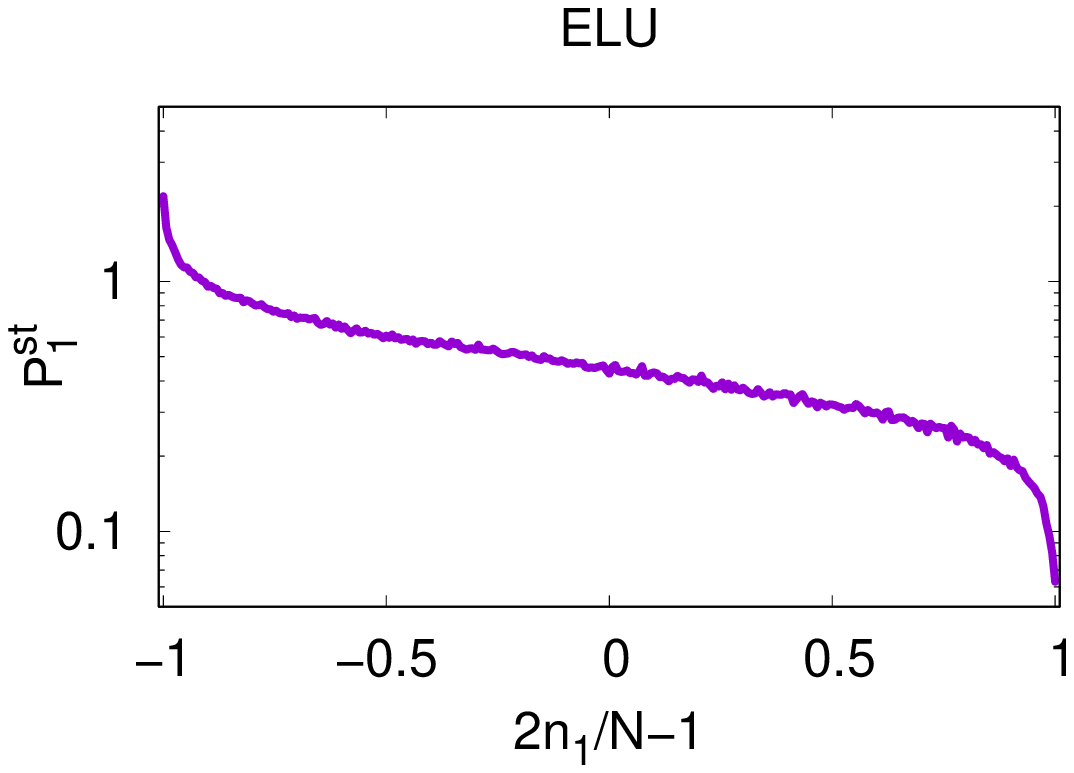}
  \includegraphics[width=.235\textwidth]{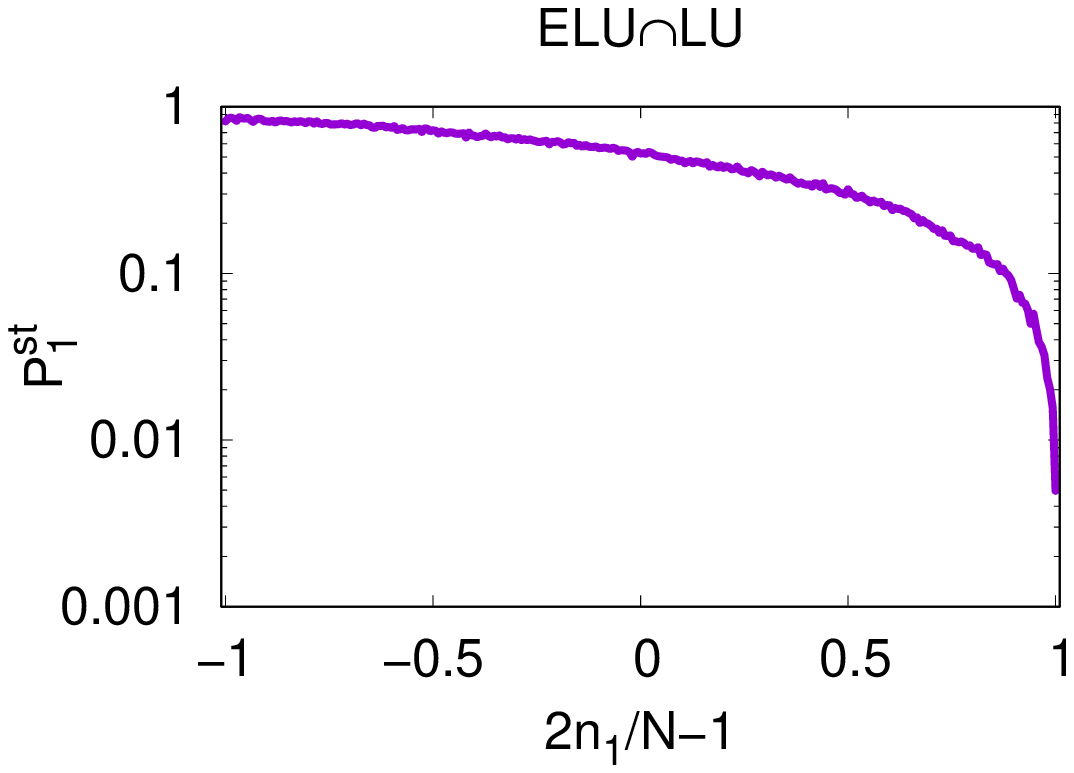}
  \includegraphics[width=.235\textwidth]{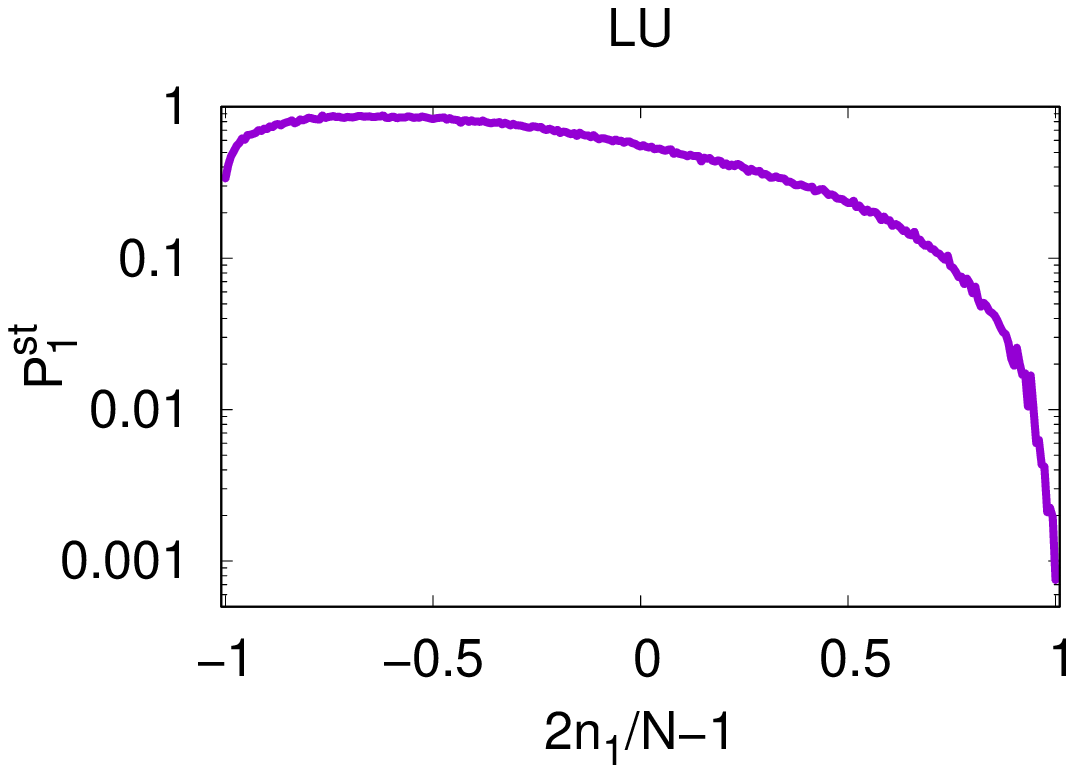}
  \caption{Marginal probability distributions for opinion $i=1$, $P_1^{\rm st}$, as a function of $2n_1/N-1$ for $N=300$, $K=2$, $M=3$, $N^{(1)}=30$, and $z_1^{(1)}=1$, all other $z_i^{(k)}=0$. From top-left to bottom-right: $\varepsilon=3.45\times 10^{-4}$ (LB$\cap$RB),  $10^{-3}$ (LB), $1.65\times 10^{-3}$ (LB$\cap$ELU), $2.25\times 10^{-3}$ (ELU), $3\times 10^{-3}$ (ELU$\cap$LU), $4\times 10^{-3}$ (LU).  The notation $\mbox{A}\cap\mbox{B}$ means points in parameter space on the line separating phases A and B. The phases indicated above each panel are those predicted by the theory for the given set of model parameters. \label{marginals_sim1}}
 
\end{figure}
We first verify the predictions for the shape of the marginal distributions. For a population of $N=300$ free agents divided into $K=2$ communities holding $M=3$ possible opinions, we take the simplest case of one zealot of opinion $1$ influencing the first community, and no other zealots ($z_1^{(1)}=1$ and $z_i^{(k)}=0$ for all other combinations of $k$ and $i$). Since $z_1=1>Z/2=1/2$, the approximate theory in Sec.~ \ref{subpopulation} predicts a phase diagram of the form shown in Fig.~\ref{phasediagrampartinf2}. In particular we expect all six phases to be physically possible for the right choice of the model parameters.

\begin{figure}[t!]
  \centering
  \includegraphics[width=.235\textwidth]{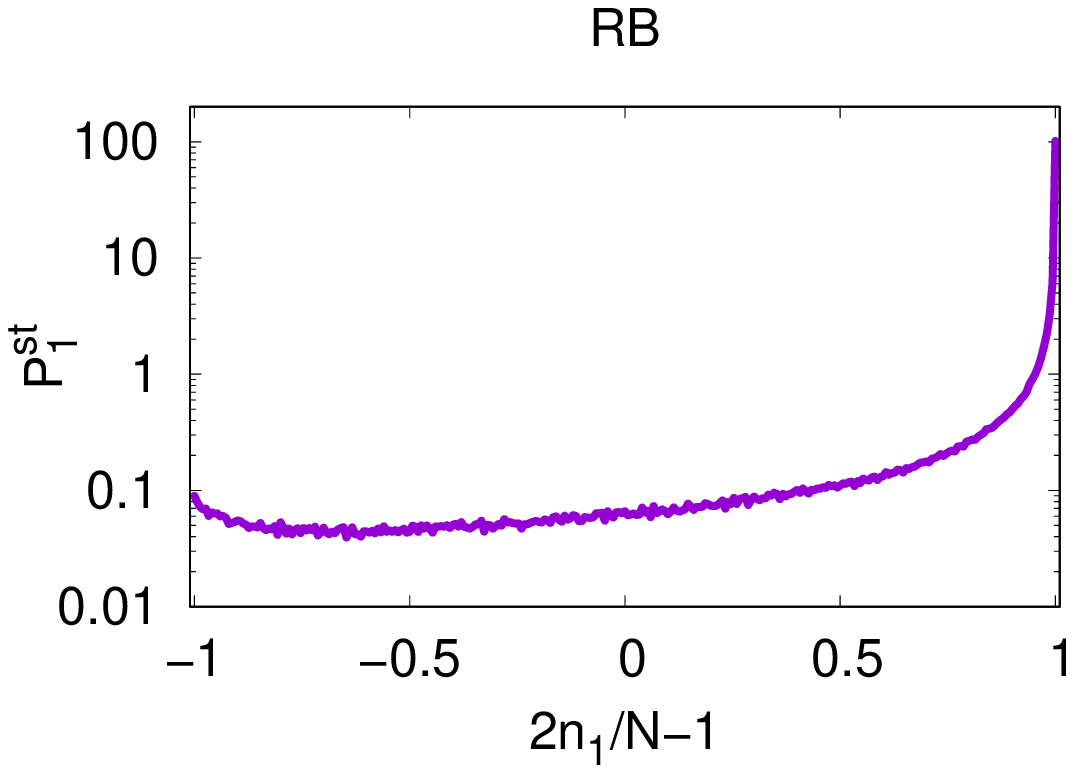}
  \includegraphics[width=.235\textwidth]{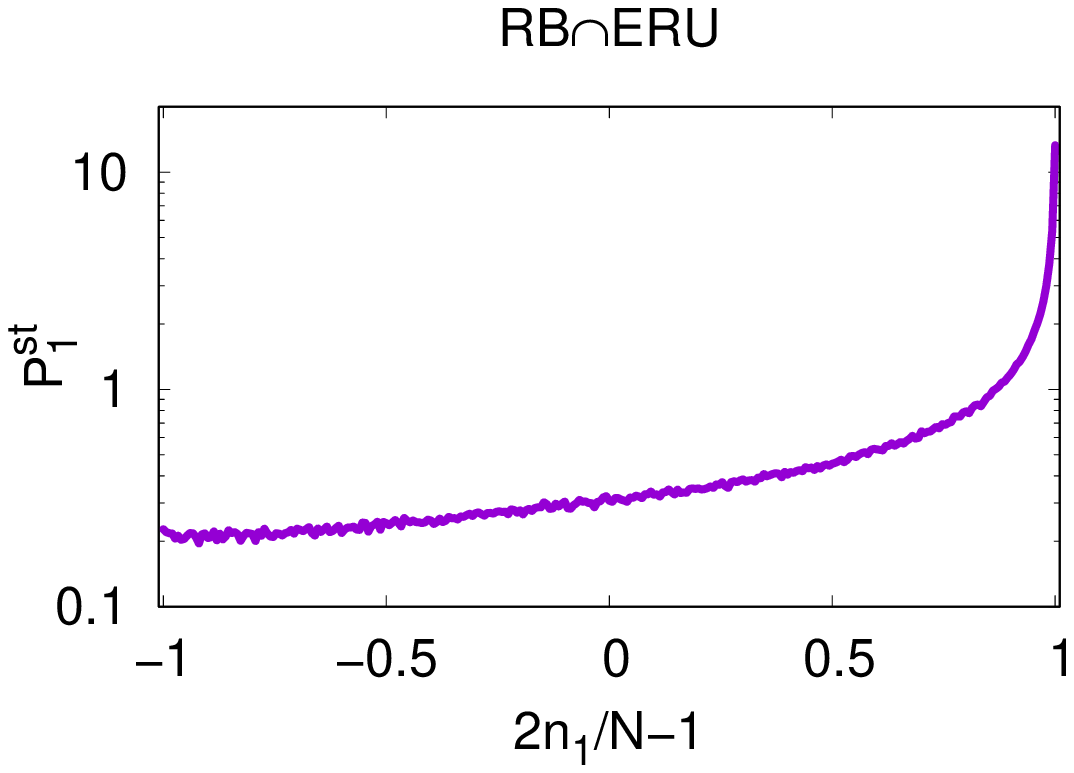}
  \includegraphics[width=.235\textwidth]{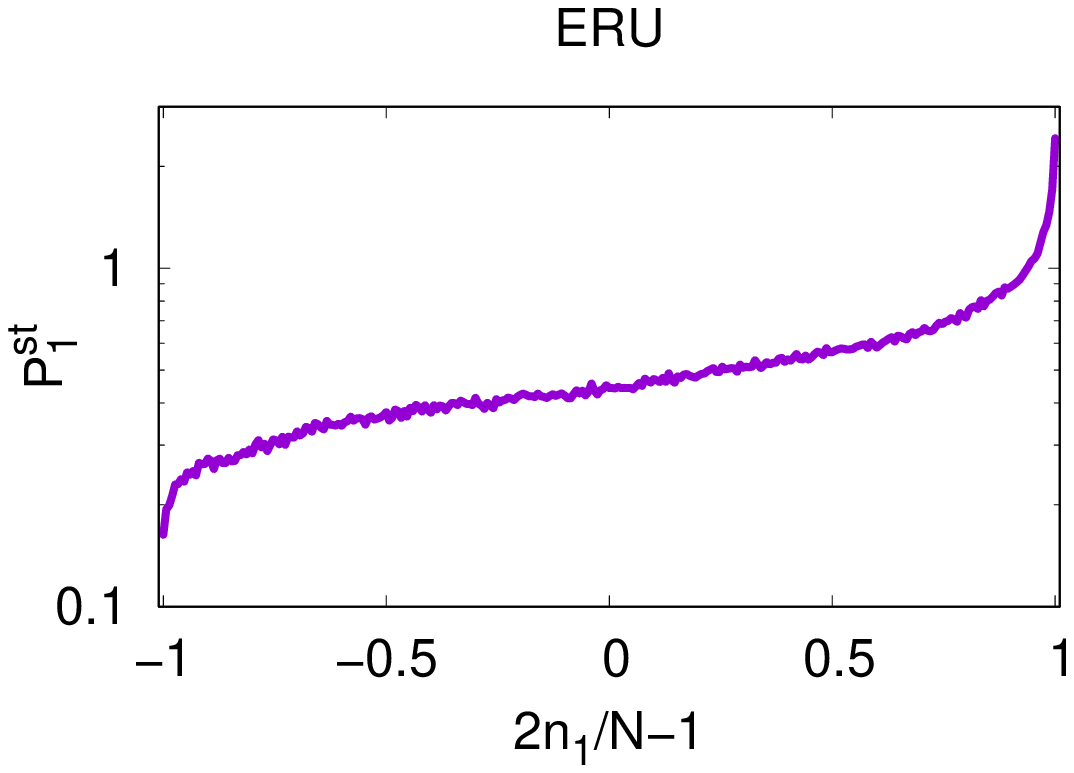}
  \includegraphics[width=.235\textwidth]{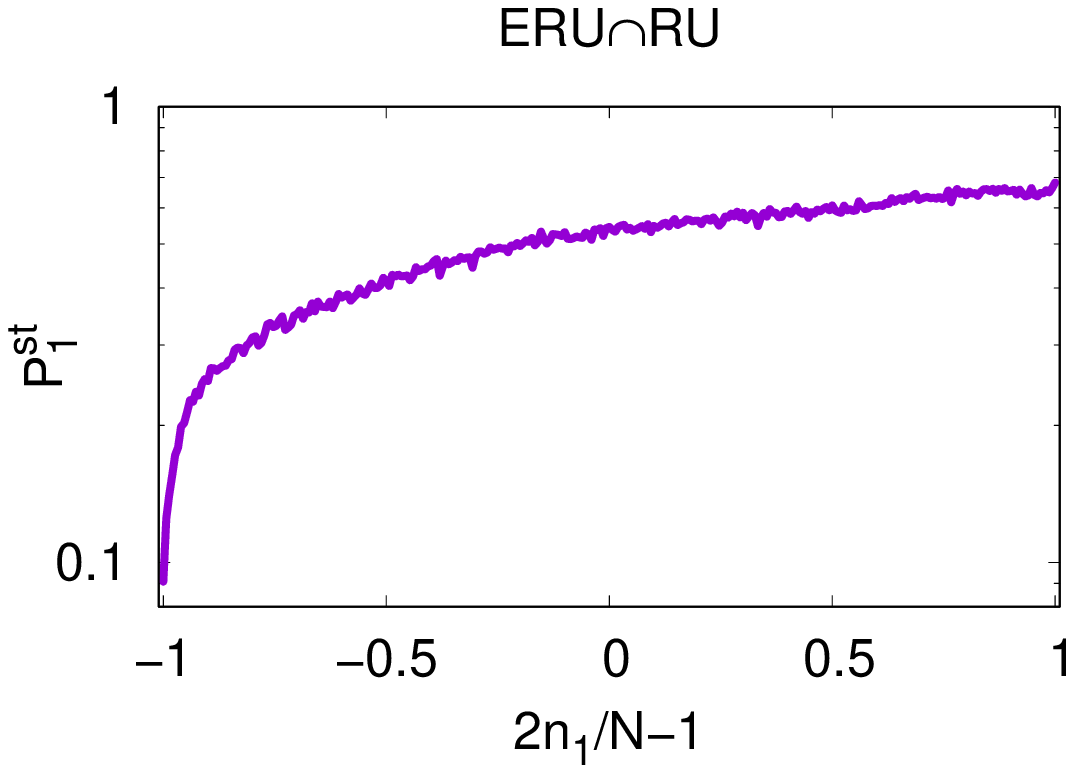}
  \includegraphics[width=.235\textwidth]{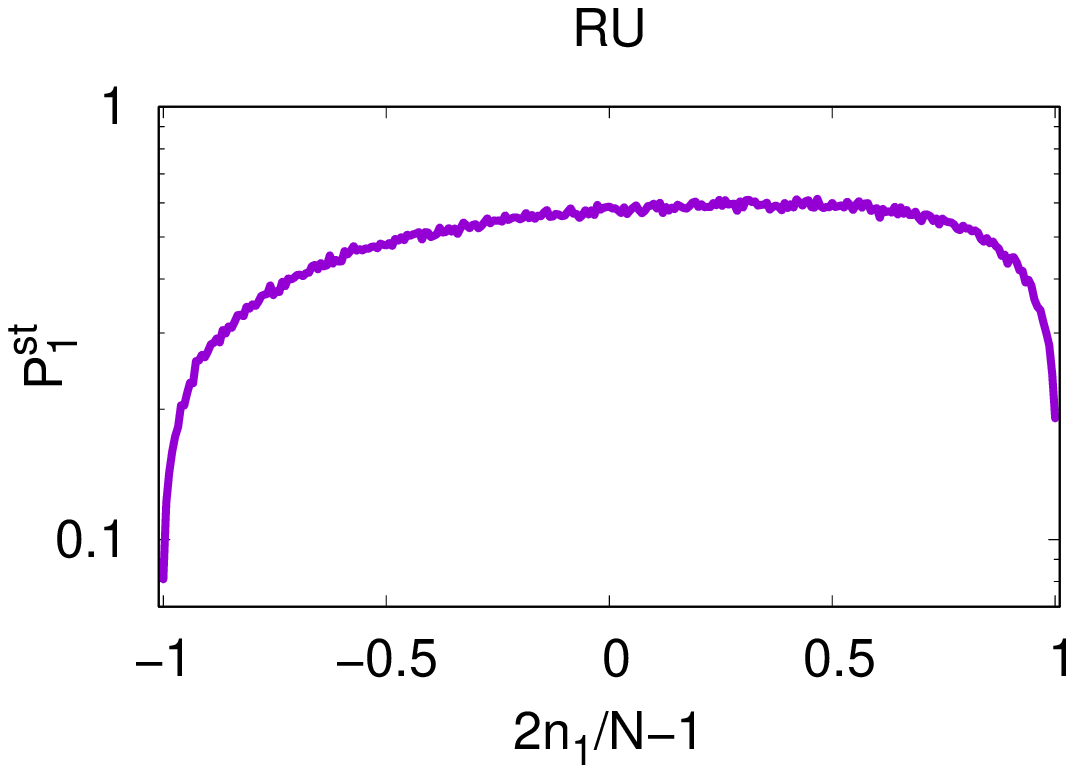}
  \includegraphics[width=.235\textwidth]{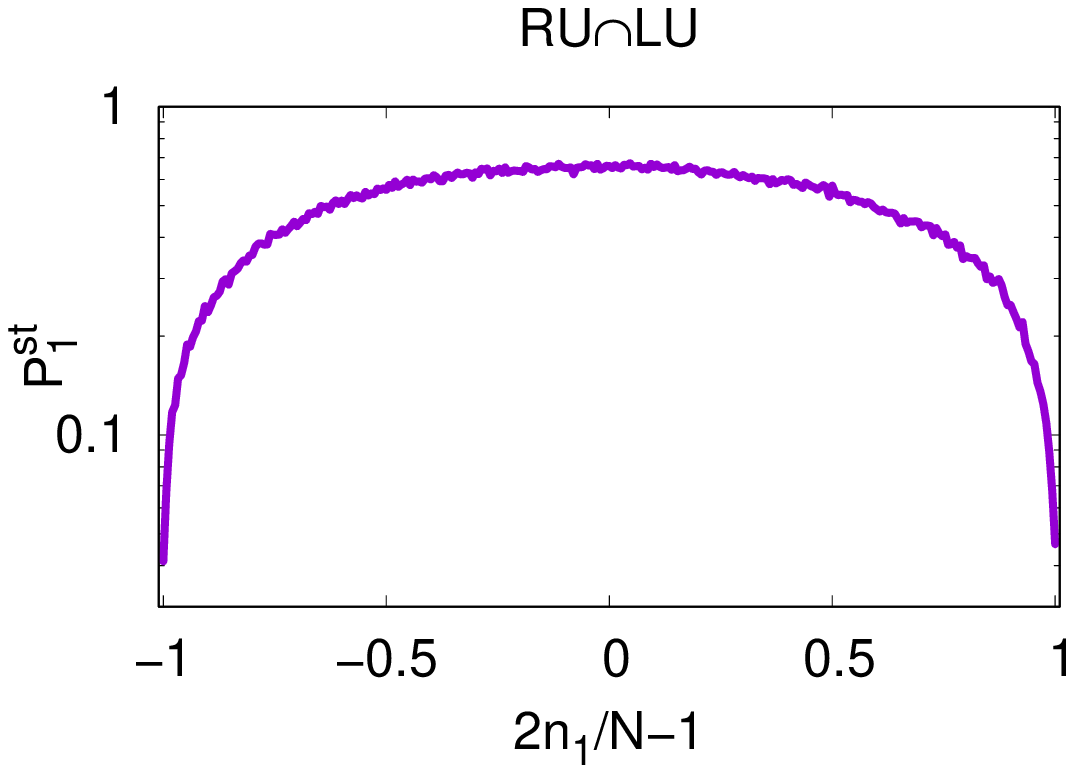}
  \caption{Marginal probability distributions for opinion $i=1$, $P_1^{\rm st}$, as a function of $2n_1/N-1$ for $N=300$, $K=2$, $M=3$, $N^{(1)}=240$, and $z_1^{(1)}=1$, all other $z_i^{(k)}=0$. From top-left to bottom-right: $\varepsilon=10^{-4}$ (RB), $6.7\times 10^{-4}$ (RB$\cap$ERU),  $1.2\times 10^{-3}$ (ERU), $1.67\times 10^{-3}$ (ERU$\cap$RU), $2.1\times 10^{-3}$ (RU), $2.65\times 10^{-3}$ (RU $\cap$LU). The phases indicated above each panel are those predicted by the theory for the given set of model parameters.\label{marginals_sim2}}
 
\end{figure}

Results from simulations are shown Figs.~ \ref{marginals_sim1} and \ref{marginals_sim2}. The two figures show the distribution of $n_1$, the total number of agents holding opinion $i=1$ across the two communities. We focus on how the marginal changes shape as we increase the mutation rate $\varepsilon$. Each panel in the two figures is labelled to indicate the shape of the marginal as predicted from the analytical approach. We use the notation $A\cap B$ to indicate the border between phases $A$ and $B$. For example, in the top left panel of Fig.~\ref{marginals_sim1} parameters are such that the system is on the line separating the LB and RB phases. The marginal is then predicted to be symmetric, with maxima of equal height at $n_1=0$ and $n_1=N$. In the panel labelled $\mbox{LB}\cap\mbox{ELU}$ the system is at the interface of the LB and ELU phases. This means that the theory predicts a single maximum at $n_1=0$, and that $P_1^{\rm st}(n_1)$ is decreasing in $n_1$, with $P_1^{\rm st}(n_1=N-1)=P_1^{\rm st}(n_1=N)$, i.e., vanishing slope at the right edge.

In Fig.~\ref{marginals_sim1} we have $N^{(1)}=30$, that is the zealots in community $k=1$ only interact directly with ten percent of all $N=300$ free agents. In Fig.~\ref{marginals_sim2} we choose $N{(1)}=240$ so that four fifths of all free agents can directly interact with zealots. The direct influence of the zealots is therefore much stronger in the second example than in the first.

For $N^{(1)}=30$, the system can be in the RB, LB, ELU, or LU phases. We show examples for LB, ELU and LU in Fig.~\ref{marginals_sim1}, as well as limiting shapes when parameters are such that the system is precisely on a line separating two phases. For $N^{(1)}=240$, the marginal can be of the RB, ERU, RU, or LU shapes, and the corresponding intermediate shapes right on the phase lines. Examples are shown in Fig.~\ref{marginals_sim2}. In all panels of both figures the shapes of the distributions obtained in simulations are as predicted from the theory, hence confirming the validity of the analytical approach, and in particular of the approximation made in Eq.~(\ref{eq:approx}).

\begin{figure}[!h]
  \centering
  \includegraphics[width=.45\textwidth]{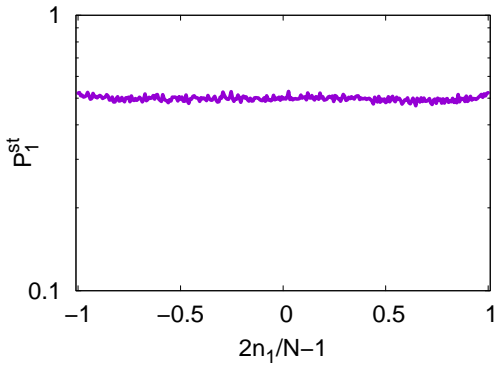}
  \caption{Marginal probability distribution for opinion $i=1$, as a function of $2n_1/N-1$ at the intersection point of the phase lines, $\varepsilon=\varepsilon_{1,r}=\varepsilon_{1,\ell}=\varepsilon_{1,c}\approx 1.65\times 10^{-3}$. Remaining parameters: $N=300$, $K=2$, $M=3$, $N^{(1)}=150$, and $z_1^{(1)}=1$, $z_i^{(k)}=0$ for all other combinations of $i$ and $k$.\label{3critical}}
\end{figure}

When the noise strength $\varepsilon$ is sufficiently small for the imitation process to dominate, the distribution for $n_1$ accumulates around extreme values of $n_1$. This is the case both in Fig.~\ref{marginals_sim1} and \ref{marginals_sim2}. For increasing mutation rate, the mode of the distribution moves to intermediate values of $n_1$, similar to what was observed for the two-opinion case \cite{nagi}. When probability accumulates near the edges then the strength of the influence of zealots determines the edge of the interval ($n_1=0$ or $n_1=N$) dominating the accumulation. When the influence of zealots is small (Fig.~\ref{marginals_sim1}) the probability tends to accumulate around smaller values of $n_1$, similar to the case without zealots $Z=0$ \cite{francisco}. On the other hand when zealots have direct influence on a larger proportion of the population (Fig.~\ref{marginals_sim2}), then probability accumulates around larger values of $n_1$. For $\varepsilon\approx  1.65\times 10^{-3}$ and $N^{(1)}=150$, the system is precisely at the intersection of all three phase lines, and the resulting marginal is flat, as shown in Fig.~\ref{3critical}.

\subsection{Phase lines}
We have conducted further tests of the analytical approximation, focusing on the quantitative verification of the phase lines. We use the  $N^{(1)}$ and $\varepsilon$ as the main control parameters. 

The different shapes of the marginals (i.e., the different phases) are identified by fixing a value of $N^{(1)}$ in simulations, and then varying $\varepsilon$. The boundaries of the phases are then found by determining the approximate values of $\varepsilon$ at which the marginal changes shape. 
 
\begin{figure*}[t]
  \centering
  \includegraphics[width=.45\textwidth]{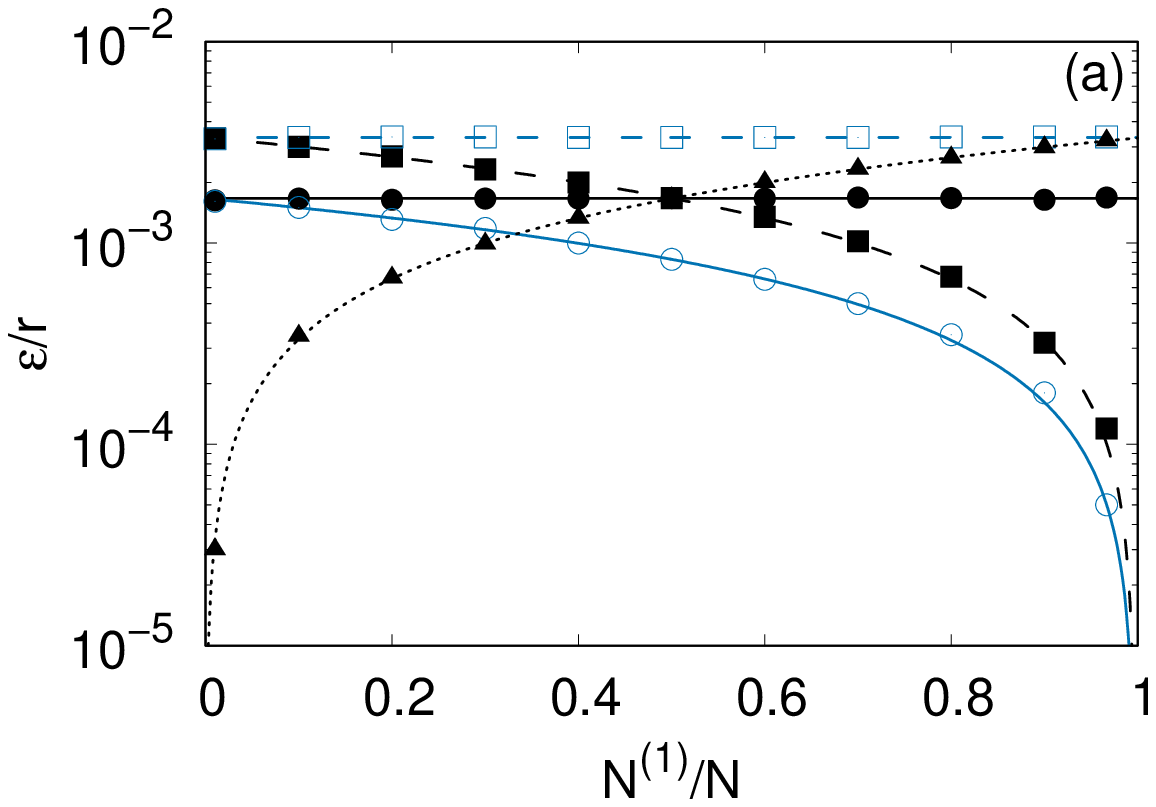}
  \includegraphics[width=.45\textwidth]{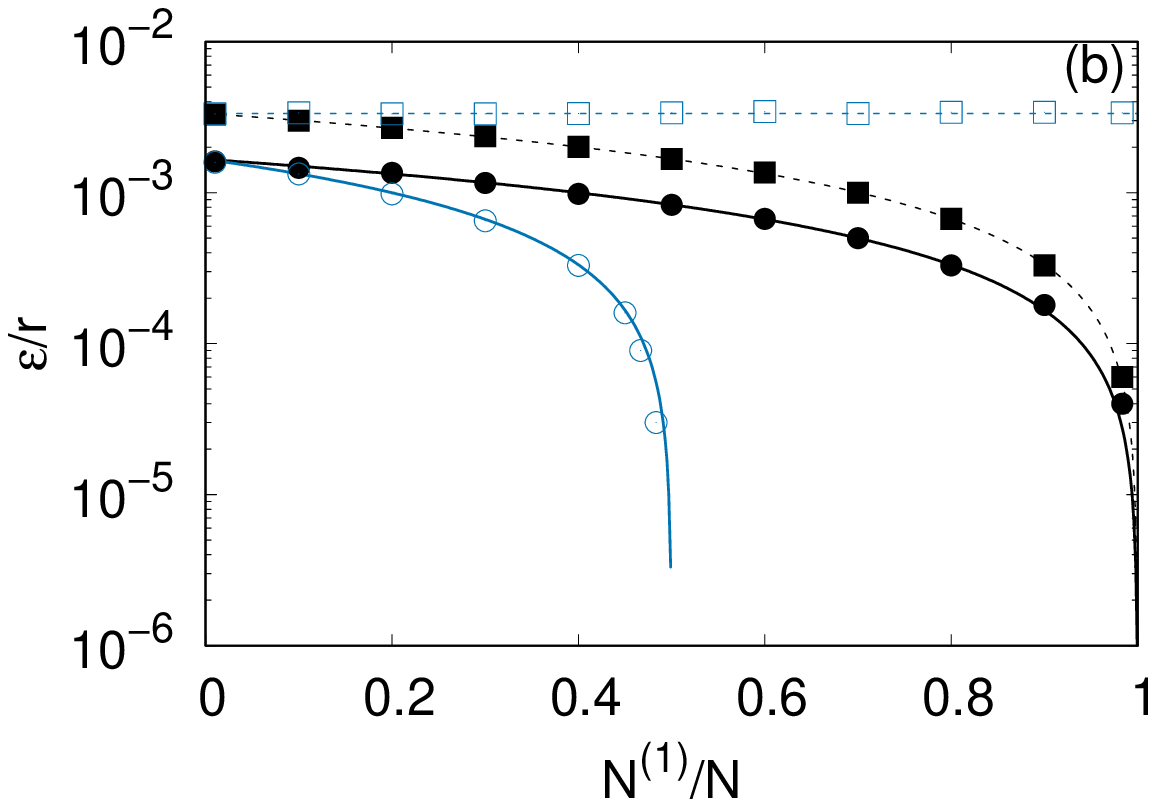}
  \caption{Phase boundaries $\varepsilon_{1,r}$ (solid lines, circles) and $\varepsilon_{1,\ell}$  (dashed line, squares) and $\varepsilon_{c,i}$ (dotted line, triangles) for $N=300$, $K=2$, and $M=3$. Panel (a): $z_1=1,\ z_2=z_3=0$; filled symbols are for opinion $1$, open symbols for opinions $2$ and $3$. Panel (b): $z_1=z_2=1,\ z_3=0$; filled symbols are for opinions $1$ and $2$, open symbols for opinion $3$. Lines are from the approximate theory of Section \ref{subpopulation} and symbols from numerical simulations of the full model defined by Eqs.~(\ref{eq:rates}).  \label{phase1}}
\end{figure*}

We start by looking at the parameters used in Figs.~\ref{marginals_sim1} and \ref{marginals_sim2}. Results for the phase lines are shown in Fig.~\ref{phase1}, for further details see also Appendix \ref{appen:2}. As seen in the figure, we find near perfect agreement despite the approximations made in the analytical approach. In panel (a) we have $z_1>Z/2$, and all six phases discussed in Fig.~\ref{figt:01} are realised for opinion $i=1$. There are no zealots for opinion states $i=2$ and $i=3$. The marginals for these two opinion states are identical by construction, but their shape differs from that for $i=1$. Given that $z_2=z_3<Z/2$ only three phases are found for the marginals of opinion states $i=2$ and $i=3$ (LU, ELU and LB).

\begin{figure*}[t]
  \centering
  \includegraphics[width=.45\textwidth]{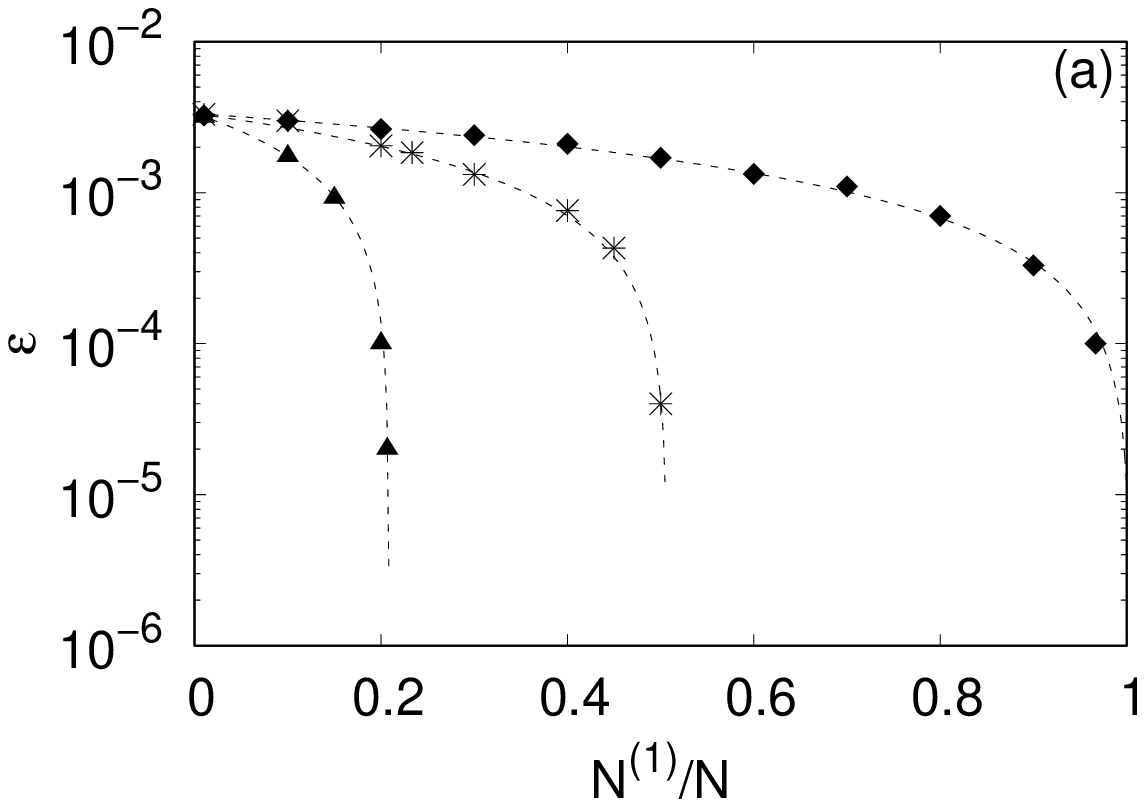}
  \includegraphics[width=.45\textwidth]{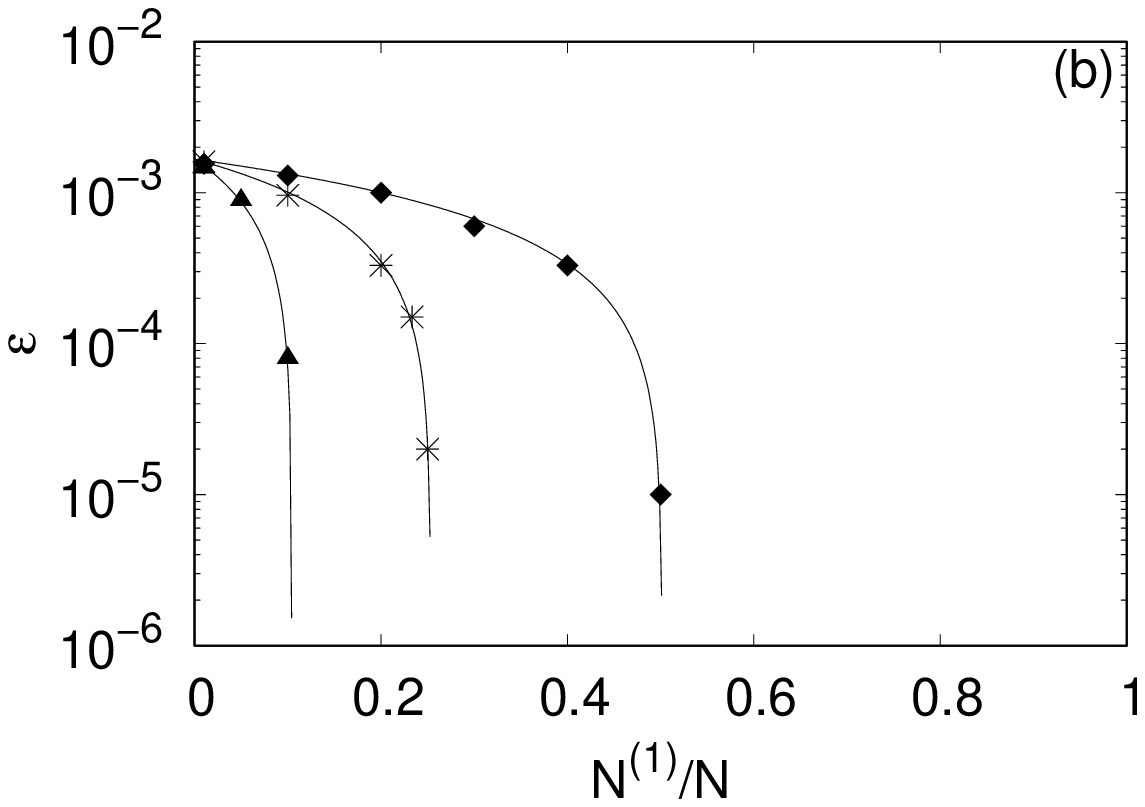}
  \caption{Phase boundaries $\varepsilon_{1,\ell}$ (a) and $\varepsilon_{1,r}$ (b) for $N=300$, $K=2$, $M=3$, $z_1=z_2=z_3$ and $z_1=1$ (diamonds), $z_1=2$ (asterisks), and $z_1=5$ (triangles). Lines are from the approximate theory, and symbols from simulations. \label{phase2}}
\end{figure*}

In Fig.~\ref{phase1}(b) we consider the case  $z_1=z_2=1$ and $z_3=0$. As before, all zealots are in community $k=1$, and there are no zealots directly affecting community $k=2$. The marginals for states $i=1$ and $i=2$ are now identical, but may differ from that for $i=3$. However, since $z_i<Z/2=1$ for all $i$, only the LU, ELU and LB phases are found. As in panel (a) numerical simulations quantitatively confirm the analytical predictions for the phase lines. As a further test we consider the case $z_1=z_2=z_3$ in Fig.~\ref{phase2}. The phase lines are then identical for the three different opinion states. Again, simulations confirm the validity of the theoretical approach.

\section{Summary}
\label{summary}

In this work we have studied the influence of zealots on the dynamics of multi-state noisy voter models with all-to-all interaction graphs. To do this we have used analytical approaches, confirmed by numerical simulations. Individuals can change states following two different mechanisms: they can copy the state of other agents, or they can change state spontaneously. Zealos are agents who can influence other agents, but who never change opinion themselves. We have considered models describing one single population of free agents and zealots, and generalisations in which there are multiple communities of free agents, each influenced by a different group of zealots. The model is motivated by processes of opinion formation, but at the same time we think its study contributes to better understanding of the effects of disorder on spin models and non-equilibrium phenomena. 

We have used the shape of the marginal stationary probability function for the number of agents with a given opinion to characterise the system. For the model with one single community we find  up to six possible phases (Fig.~\ref{figt:01}) as the noise strength and the number of zealots varied. However, not all of these phases can be realised at physically meaningful parameters. Our work generalises findings from existing studies of multi-state noisy voter models without zealots \cite{francisco}, and of two-state noisy voter models with zealots \cite{nagi}. 

Symmetric noisy two-state voter models show a transition between a state with a bimodal stationary distribution and a state with unimodal stationary distribution \cite{nvm1,nvm2,nvm3,nvm4,nvm5,nvm6,nvm7,nvm8} . One main result of \cite{francisco} is the observation that this transition splits up into sequences of different transitions at the right and left edges of marginal stationary distributions in multi-state noisy voter models.  In \cite{nagi} it was found that the presence of zealots can remove unimodal behaviour in two-state models. The analysis in this paper shows that these statements transfer to multi-state noisy voter models with zealots. We observe separate changes of shape at the right and left edges of the marginals of the stationary distribution, indicating that there is no single transition between a unimodal and a multimodal state. At the same time, the presence of zealots can remove the transition at the right edge. In contrast to the two-state model we find that shape-changes at the left egde are possible even in the presence of zealots. Contrary to the multi-state model without zealots, flat marginal distributions are possible for selected model parameters (Fig.~\ref{3critical}).

The analysis can be extended to compartmental models, in which the population of free agents divides into several communities. We have shown that analytical progress is possible for such a model, based on the approximation in Eq.~(\ref{eq:approx}). Numerical simulations confirm that the resulting predictions are qualitatively and quantitatively accurate. 

Using the approximation an interesting connection between the model with multiple communities and an effective single-community model emerges. Multi-community models can be mapped onto a single-community model with an effective imitation rate and a non-integer number of zealots.  Alternatively, the effective dynamics can be interpreted as a model with an integer number of `soft' zealots, who are able to change the states of free agents not with certainty upon interaction, but only with a certain probability. This can be seen from Eq.~(\ref{eq:bdrates_allk}), and is discussed in more detail in Appendix \ref{app2_1}. 

Further, the model with zealots can be mapped onto a noisy voter model with heterogeneous mutation rates, similar to the one studied in \cite{francisco}. This can be seen from Eqs.~(\ref{eq:bdratesall}), which can be re-written as
\BE
T_i^+&=&\frac{(N-n_i)n_i}{N+Z}+\sum_{j\neq i}\varepsilon_{j\to i} n_j, \nonumber \\
T_i^-&=&\frac{n_i(N-n_i)}{N+Z} +\sum_{j\neq i} \varepsilon_{i\to j} n_i,
\EE
with $\varepsilon_{i\to j}=\varepsilon+z_j/(N+Z)$. A similar mapping can be performed starting from Eqs.~(\ref{eq:bdrates_allk}).

The results of our work are not restricted to the multi-state noisy voter models with zealots, but include a family of models whose rates can be written or approximated as those of a birth-death process of the form $T^\pm (n)=a^\pm+b^+n+cn^2$, with $a^\pm\ge 0$, $b^\pm$, $c$ so that $T^\pm\ge 0$ for any possible $n$. In particular, we would expect models in this class to have similar phases as the ones for the current model.

As a final note, we remark that it is not entirely obvious when the approximation in Eq.~(\ref{eq:approx}) is valid. The fact that the outcomes of our analytical work are confirmed in simulations suggests that approaches based on approximations of this type can be useful for compartmental individual-based models in other contexts. Further investigation is needed to understand the nature and validity of the approximation.

\acknowledgments
TG acknowledges funding from the Spanish Ministry of Science, Innovation and Universities, the Agency AEI and FEDER (EU) under the grant PACSS (RTI2018-093732-B-C22), and the Maria de Maeztu program for Units of Excellence in R\&D (MDM-2017-0711).

\appendix

\section{Proof of the lemmas}
\label{appen:1}
\subsection{Proof of Lemma 1}\label{sec:proof1}
We prove the following result, which is a generalization of Lemma 1 of the main text: \\

\emph{Proposition 1}: The steady-state probability distribution $P^{\rm st}(n)$ of a one-step Markov process for a discrete variable $n\in\{0,1,\dots,N\}$ with rates
\begin{eqnarray}
  \label{eq:aprpm}
  T^+(n)=a^++b^+n+cn^2, \nonumber \\
  T^-(n)=a^-+b^-n+cn^2,
\end{eqnarray}
has at most one extremum in $n=1,\dots,N-1$.

\emph{Proof}. Suppose $n_M\in\{1,\dots,N-1\}$ is a maximum of $P(n)$. Then $P^{\rm st}(n_M-1)< P^{\rm st}(n_M)$, and $P^{\rm st}(n_M)> P^{\rm st}(n_M+1)$. Using the steady-state condition $T^+(n-1)P^{\rm st}(n-1)=T^-(n)P^{\rm st}(n)$, the latter inequalities turn into $T^-(n_M)< T^+(n_M-1)$ and $T^-(n_M+1)> T^+(n_M)$. Using the explicit form of the rates, this is equivalent to the conditions 
\begin{equation}\label{eq:BA}
  B-A< A n_M< B,
\end{equation}
with $A\equiv b^--b^++2c$ and $B\equiv a^+-a^--b^++c$. Division by $A$ (and inverting the inequality signs as appropriate, depending on the sign of $A$) shows that there is at most one possible natural number $n_M$ fulfilling both inequalities in Eq.~(\ref{eq:BA}) .

Analogously, in order for $n_m$ to be a minimum of $P(n)$, we require
\begin{equation}\label{eq:BA2}
  B-A> A n_m> B,
\end{equation}
which, again, only has at most one possible integer solution $n_m$.

For a given set of parameters, $A$ and $B$ are fixed. There can then not be simultaneous solutions $n_M$ and $n_m$ of the relations in Eq.~(\ref{eq:BA}) and (\ref{eq:BA2}) respectively, as this would require $B-A<B$ and $B-A>B$ at the same time. Hence, $P^{\rm st}(n)$ can have at most one extremum in $\{1,\dots,N-1\}$. $\blacksquare$ 

Lemma 1 of the main text follows after realizing that the rates in Eqs.~\eqref{eq:bdratesall} can be written in the form in Eqs.~\eqref{eq:aprpm}. 

\subsection{Proof of statement related to Lemma 2}\label{sec:proof2}
Assume $N$ is even, and that the marginal for opinion state $i$ has a maximum at $n_i=N/2$. Following the argument in Appendix \ref{sec:proof1} this requires $T^+_i(N/2-1)>T_i^-(N/2)$ and $T_i^-(N/2+1)>T_i^+(N/2)$. Using the rates in Eq.~(\ref{eq:bdratesall}), the first of these conditions translates into
\be\label{eq:aux1}
\varepsilon<\frac{z_i-Z_{-i}-2(1-z_i)/N}{(N+Z)(M-2-2/N)}.
\ee

The condition $T_i^-(N/2+1)>T_i^+(N/2)$ on the other hand turns into 
\be\label{eq:aux2}
\varepsilon>\frac{z_i-Z_{-i}+2(1-Z_{-i})/N}{(N+Z)[M-2+2(M-1)/N]}.
\ee
The expression on the right-hand side in (\ref{eq:aux1}) is larger than $\varepsilon_{c,i}$ in Eq.~(\ref{eq:eps_c}), and that on the right-hand side of (\ref{eq:aux2}) is smaller. This means that the distribution $P_i^{\rm st}$ attains its maximum at $n_i=N/2$ in a corridor of values for $\varepsilon$. This is a natural consequence of $n_i$ being a discrete variable.  As $\varepsilon$ is varied, the location of the maximum of $P_i^{\rm st}$ jumps from one integer value of $n_i$ to the next, but remains fixed at any one value of $n_i$ throughout a finite interval of values for $\varepsilon$. The corridor includes $\varepsilon_{c,i}$ and its width is proportional to $1/N$. 

Analogous arguments can, in principle, be developed for the case of a minimum at $n_i=N/2$. Such a minimum would be realised on the segment of the line $\varepsilon=\varepsilon_{c,i}$ separating the LB and RB phases in Fig.~\ref{figt:01}. We notice however that this segment cannot physically be reached for $M>2$, see the discussion in Sec.~\ref{sec:unphysical}.

\subsection{Proof of Lemma 3}
We prove the following, more general statement:
\medskip

\emph{Proposition 3}: The steady-state probability function $P^{\rm st}(n)$ of a one-step Markov process of a discrete variable $n\in\{0,1,\dots,N\}$, with rates given by Eqs.~\eqref{eq:aprpm} is flat if and only if $P(n)=P(n-1)$ for at least two different values of $n\in\{1,\dots,N\}$.

\medskip

\emph{Proof}. If the distribution is flat, then $P^{\rm st}(n)=P^{\rm st}(n-1)$ for all $n=1,\dots, N$ so the condition of the proposition is fulfilled. To prove the reverse, we note that the equality $P^{\rm st}(n-1)=P^{\rm st}(n)$ is equivalent to the condition $T_i^+(n-1)=T_i^-(n)$, due to the fact that the equality $P^{\rm st}(n-1)T_i^+(n-1)=P^{\rm st}(n)T_i^-(n)$ holds at stationarity. Using the rates in Eqs.~\eqref{eq:aprpm}, last condition turns into a linear equation for $n$ of the form $C+Dn=0$, where $C$ and $D$ are functions of the parameters of the system, but independent of $n$. If the condition $C+Dn=0$ holds for at least two different values of $n$, then necessarily $C=D=0$. The condition then holds for all $n$. Hence, $P^{\rm st}(n)$ is flat. $\blacksquare$ 

\medskip

Lemma 3 in the main text follows from the fact that $P_i^{\rm st}(n_i-1)=P_i^{\rm st}(n_i)$ holds for $n_i=1$ and $n_i=N$ at the intersection point of the lines $\varepsilon_{\ell,i}$ and $\varepsilon_{r,i}$. 

\section{Further properties of the model with $K=2$ and $Z^{(2)}=0$}
\label{appen:2}
\subsection{Discussion and interpretation of the model}\label{app2_1}
If there are no zealots in community $k=2$, then $Z=Z^{(1)}$. As a consequence, the weights $q\kk$ in Eq.~(\ref{eq:qk}) reduce to
\be\label{eq:qq}
q^{(1)}=\frac{N^{(1)}}{N}, ~q^{(2)}=\left(1-\frac{N^{(1)}}{N}\right)\frac{N+Z}{N}.
\ee
Writing further 
\be
x^{(1)}\equiv \frac{N^{(1)}}{N}, ~x^{(2)}\equiv\frac{N^{(2)}}{N},
\ee
we also have
\be
\tilde z_i =x^{(1)} z_i, ~\tilde Z_{-i}=x^{(1)} Z_{-i},
\ee
and $\alpha=x^{(1)}+x^{(2)}\frac{N+Z}{N}$. The transition rates in Eqs.~(\ref{eq:bdrates_allk}) then become

\begin{eqnarray}\label{eq:aux6}
  T_i^+(n_i)&=& (N-n_i)\left[x^{(1)}\frac{n_i+z_i}{N+Z}+x^{(2)}\frac{n_i}{N}\right]+\varepsilon(N-n_i), \nonumber \\
  T_i^-(n_i)&=&n_i \left[x^{(1)}\frac{(N-n_i)+Z_{-i}}{N+Z}+x^{(2)}\frac{N-n_i}{N}\right] \nonumber \\
  &&+(M-1)\varepsilon n_i, \nonumber \\
\end{eqnarray}
These rates have a direct physical interpretation. We illustrate this for the rate $T_i^+$. The term proportional to $\varepsilon$ describes mutation, and is independent of the community structure. The first term can be thought of as follows: A free agent is chosen at random from the entire population for potential adoption of state $i$. This only contributes to $T_i^+$ if this agent is not already of type $i$, hence the factor $N-n_i$. We will refer to this individual as `agent 1'. Given the assumption $n_i^{(1)}/N^{(1)}=n_i^{(2)}/N^{(2)}$, the probability that agent 1 is from community $k=1$ is $x^{(1)}$, and that for being drawn from community $k=2$ is $x^{(2)}$ ($x^{(1)}+x^{(2)}=1$). If agent 1 is from community $k=1$, then an interaction partner (agent 2) is chosen at random from the pool all individuals agent 1 can interact with. This pool consists of all $N$ free agents and all $Z=Z^{(1)}$ zealots. The probability that agent 2 is of type $i$ is then $(n_i+z_i)/(N+Z)$. If however, agent 1 is from community $k=2$, then interaction is only possible with one of the $N$ free agents. The probability that the interaction partner is of type $i$ is then $n_i/N$. 

Alternatively, we can write the rate $T_i^+$ in Eq.~(\ref{eq:aux6}) in the following form
\BE \label{eq:aux7}
T_i^+(n_i)&=& \frac{(N-n_i)n_i}{N+Z} \left[x^{(1)}+\frac{N+Z}{N}x^{(2)}\right]\nonumber \\
&&+\frac{N-n_i}{N+Z} x^{(1)} z_i \nonumber \\
&&+\varepsilon(N-n_i),
\EE

The first term describes interaction between two free agents. Comparing this expression to that in Eq.~(\ref{eq:bdratesall}) for the single-community model we note the factor $x^{(1)}+\frac{N+Z}{N}x^{(2)}>1$. This enhancement of the rate with which pairs of free agents interact arises from the fact that free agents in community $k=2$ are guaranteed to interact with a second free agent once chosen for interaction. The expression in Eq.~(\ref{eq:bdratesall}) on the contrary is for a single-population model in which a free agent, once chosen for interaction, interacts with a second free agent only with probability $N/(N+Z)$.

The second term on the right-hand side of Eq.~(\ref{eq:aux7}) can be interpreted as interaction with a `soft zealot'. A free agent who is not of type $i$ is chosen for potential update, and then interacts with a zealot of type $i$. That zealot however only manages to change the free agent's state with probability $x^{(1)}$. 

The third term in Eq.~(\ref{eq:aux7}) finally describes spontaneous opinion changes as before.
\subsection{Right edge}
The expression for $\varepsilon_{r,i}$ is given by
\be\label{eq:aux4}
\varepsilon_{r,i}=\frac{\alpha}{N+Z}\frac{\tilde z_i/\alpha - 1 - N(\tilde Z_{-i}/\alpha-1)}{(M-1)N-1}.
\ee
 
For $N^{(1)}=0$ the model reduces to the case of a single community (community $k=2$) with no zealots ($\alpha=1$, $\tilde z_i=\tilde Z_{-i}=0$). We then find 
\begin{equation}\label{eq:aux3}
  \varepsilon_{r,i}=\frac{N-1}{N[(M-1)N-1]},
\end{equation}
which is equivalent to Eq. (16) in \cite{francisco}. Assuming $z_i\ne \frac{N^2+N-1}{N(N+1)} Z$, we have $\varepsilon_{r,i}=0$ when
\begin{equation}
  \frac{N^{(1)}}{N}=\frac{(N-1)(N+Z)}{(N^2+N-1)Z-N(N+1)z_i}. 
\end{equation}
Given that $\tilde Z_{-i}=\tilde Z-\tilde z_{i}$, Eq.~(\ref{eq:aux4}) shows that the values $\varepsilon_{r,i}$ and $\varepsilon_{r,j}$ for two different opinion states $i\neq j$ coincide when $N^{(1)}=0$ [Eq.~(\ref{eq:aux3})] or when $z_i=z_j$.

\subsection{Left edge}
For the model with two communities and no zealots in community $k=2$ we have
\be\label{eq:aux5}
\varepsilon_{\ell,i}= \frac{\alpha}{N+Z}\frac{\tilde Z_{-i}/\alpha-1-N(\tilde z_i/\alpha-1)}{N+1-M}. 
\ee
For $N^{(1)}=0$ this reduces to
\begin{equation}
  \varepsilon_{\ell,i} =\frac{1}{N}\frac{N-1}{N+1-M}, 
\end{equation}
which is equivalent to Eq. (15) of \cite{francisco}. If $z_i\ne \frac{Z}{N(N+1)}$ then $\varepsilon_{\ell,i}=0$ is equivalent to
\begin{equation}
  \frac{N^{(1)}}{N}=\frac{(N-1)(N+Z)}{N(N+1)z_i-Z}. 
\end{equation}
Similar to the right edge, $\varepsilon_{\ell,i}=\varepsilon_{\ell,j}$ for two different opinions $i\neq j$ when $N^{(1)}=0$ or when $z_i=z_j$.

Moreover, for a given opinion $i$, and $N^{(1)}=0$,
\begin{equation}
\varepsilon_{r,i}< \varepsilon_{\ell,i}.  
\end{equation}
for $M>2$. For $M=2$ (and still assuming $N^{(1)}=0$) one has $\varepsilon_{r,i}= \varepsilon_{\ell,i}$. In this latter case, the model reduces to the symmetric two-state noisy voter model without zealots. 

Focusing now on the model with $M>2$ and general values for $N^{(1)}$,  we find that $\varepsilon_{r,i}=\varepsilon_{\ell,i}$ if and only if
\begin{eqnarray}\label{eq:aux12}
  \frac{N^{(1)}}{N}&=&\frac{(N+Z)(M-2)}{NM}\frac{1}{z_i-\frac{N+2-M}{N}\frac{Z}{M}}, \nonumber\\
\end{eqnarray}
assuming $z_i>Z/2$.
If this condition is fulfilled then
\begin{eqnarray}
  \varepsilon_{r,i}=\varepsilon_{\ell,i}&=&\frac{2}{NM}\frac{z_i-\frac{Z}{2}}{z_i-\frac{N+2-M}{N}\frac{Z}{M}}.  
  \end{eqnarray}

The condition $z_i>Z/2$ ensures $z_i>(N+2-M)Z/(NM)$ for $M\ge 2$, hence $\varepsilon_{r,i}= \varepsilon_{\ell,i}>0$ and $N^{(1)}/N>0$. The conditions also ensure that $N^{(1)}/N<1$.    For $Z=1$, and assuming $z_i>Z/2$, the expression in Eq. (\ref{eq:aux12}) takes its maximum at $z_i=1$, resulting in $N^{(1)}/N =(N+1)(M-2)/[N+(N+1)(M-2)]<1$. For $Z\ge 2$ (and $z_i>Z/2$) the denominator on the right-hand side of Eq.~(\ref{eq:aux12}) takes its minimum for $z_i=Z/2$, hence $N^{(1)}/N<\frac{(N+Z)(M-2)}{NM}\frac{1}{Z/2-\frac{N+2-M}{N}\frac{Z}{M}}= \frac{2}{Z}\frac{N+Z}{N+2}$. The last expression is smaller than or equal to one for $Z\ge 2$, since it is a decreasing function of $Z$, and equal to $1$ for $Z=2$.

We note that $z_i>Z/2$ can only be fulfilled by one opinion.   
 
\subsection{Further properties of the phase lines in the limit $N\gg Z,M$}
In the limit $N\gg Z,M$ one has
\be
\alpha=q^{(1)}+q^{(2)}\approx 1,
\ee
using the relations in Eq.~(\ref{eq:qq}). From this, and $\tilde z_i=\frac{N^{(1)}}{N}z_i, \tilde Z_{-i}=\frac{N^{(1)}}{N}Z_{-i}$ (which hold whenever $Z^{(2)}=0$), we then find
\begin{eqnarray}
  &&\varepsilon_{r,i}\approx \frac{1}{(M-1)N}\left(1-\frac{N^{(1)}}{N}Z_{-i}\right) \nonumber\\
  &&\varepsilon_{\ell,i}\approx\frac{1}{N}\left(1-\frac{N^{(1)}}{N}z_i\right), \nonumber \\
  &&\varepsilon_{c,i}\approx\frac{z_i-Z_{-i}}{(M-2)N}\frac{N^{(1)}}{N}.\label{eq:aux10}
\end{eqnarray}
in the limit $N\gg Z,M$. These expressions reduce to those in Eqs. \eqref{eq:clra} when $N^{(1)}=N$.

The approximations in Eq.~\eqref{eq:aux10} for the case of two communities allow us to infer further properties of the phase lines. The dependence of $\varepsilon_{r,i}$ on $z_i$ is only through $Z_{-i}$ in the limit $N\gg Z,M$, and $\varepsilon_{r,i}$ is a decreasing function of $N^{(1)}$. When there are no zealots for any opinion $j\neq i$ (i.e., when $Z_{-i}=0$) then $\varepsilon_{r,i}$ has no dependence on $N^{(1)}$.

The dependence of $\varepsilon_{\ell,i}$ on the number of zealots is through $z_i$ in the limit $N\gg Z,M$. The value of $\varepsilon_{\ell,i}$ is a decreasing function of $N^{(1)}$, and constant for $z_i=0$ within the approximation of Eq.~(\ref{eq:aux10}).

Finally, within the approximation, the lines defined by $\varepsilon_{r,i}$ and $\varepsilon_{\ell,i}$ for a particular opinion $i$ never cross in the phase diagram, for $z_i\leq Z/2$, that is to say we always have $\varepsilon_{r,i}<\varepsilon_{\ell,i}$. To demonstrate this, we show that $\varepsilon_{r,i}=\varepsilon_{\ell,i}$ is possible only for a negative value of  $\varepsilon_{r,i}$ and $\varepsilon_{\ell,i}$:
 \begin{eqnarray}
   \nonumber 
   && \varepsilon_{r,i}=\varepsilon_{\ell,i} \Leftrightarrow \frac{1}{(M-1)N}\left[1-\frac{N^{(1)}}{N}\left(Z-z_i\right)\right]\\ \nonumber && =\frac{1}{N}\left(1-\frac{N^{(1)}}{N}z_i\right) \\ \nonumber  && \Rightarrow \frac{N^{(1)}}{N}=\frac{M-2}{(M-1)z_i-Z_{-i}}\\ \nonumber && =\frac{1}{z_i-(Z-2z_i)/(M-2)}\ge \frac{1}{z_i} \\ && \Rightarrow  \varepsilon_{r,i}=\varepsilon_{\ell,i}<0.
 \end{eqnarray}


\begin{thebibliography}{10}
\bibitem{note1} Noting the remarks in Appendix \ref{sec:proof2} an exremum at $n_i=N/2$ can still be possible in a small range of $\varepsilon$ even if $\varepsilon_{c,i}<0$, provided $N$ is sufficiently small. Once $N$ exceeds a a threshold set by the other model parameters, no extremum at $n_i=N/2$ can be attained.
\end{thebibliography}
\end{document}